\documentclass[nonacm,sigplan]{acmart}
\usepackage{natbib}
\usepackage{amsmath,amsfonts}
\usepackage{algorithm}
\usepackage{algorithmic}
\usepackage{graphicx}
\usepackage{textcomp}
\usepackage{xcolor}
\usepackage{fancyhdr}
\usepackage{hyperref}

\usepackage{subfigure}
\usepackage[braket, qm]{qcircuit}
\usepackage{mathtools}
\usepackage{ulem}

\hypersetup{
  colorlinks   = true,    
  urlcolor     = blue,    
  linkcolor    = blue,    
  citecolor    = red      
}

\title{MagicPool: Dealing with Magic State Distillation Failures on Large-Scale Fault-Tolerant Quantum Computer}

\author{Yutaka Hirano}
\affiliation{%
  \institution{Osaka University}
  \country{Japan}
}%
\author{Yasunari Suzuki}
\affiliation{%
  \institution{NTT Computer and Data Science Laboratories, NTT Corporation}
  \country{Japan}
}%
\author{Keisuke Fujii}%
\affiliation{%
  \institution{Osaka University}
  \country{Japan}
}%
\affiliation{%
  \institution{RIKEN Center for Quantum Computing (RQC)}
  \country{Japan}
}

\begin{document}
\begin{abstract}
Magic state distillation, which is a probabilistic process used to generate magic states, plays an important role in universal fault-tolerant quantum computers.
On the other hand, to solve interesting problems, we need to run complex programs on fault-tolerant quantum computers, and hence, the system needs to use hardware resources efficiently.
Taking advantage of parallelism is a major optimization strategy and compilers are responsible for performing optimizations to allow parallel processing.
However, the probabilistic nature of magic state distillation is not compatible with compile-time optimizations and results in an additional run-time delay.
To reduce the additional run-time delay, we propose introducing a pool of magic states.
We run simulations of quantum circuits to verify the magnitude of the run-time delay and the usefulness of the mitigation approach.
The experimental results show that the run-time delay is amplified by parallel processing, and pooling effectively reduces the run-time delay with a small spatial cost.
\end{abstract}

\maketitle 
\pagestyle{plain} 

\section{Introduction}

Quantum computers promise to solve interesting problems that are also believed
to be difficult for classical computers, such as quantum physics simulation \cite{Abrams1999Quantum} and
integer factoring \cite{Shor1994Factoring}.
Currently available quantum computers, also known as Noisy Intermediate-Scale Quantum computers (NISQ), are too vulnerable to hardware
errors to run complex quantum algorithms such as Shor's factoring algorithm \cite{Shor1994Factoring};
hence, to fully realize the promise we need Fault-Tolerant Quantum Computers
(FTQC) built on top of error correcting code.

An error correcting code used by FTQC suppresses errors by encoding one logical qubit
to multiple physical qubits.
Gates that operate on logical qubits also need to be constructed.
Transversal gates are fault-tolerant and easy to implement, but no error correcting code has a both universal and transversal gate set \cite{Eastin2009RestrictionsOnTransversal},
which means that a special procedure is required for universal fault-tolerant quantum computing.
One approach is to use a special quantum state called a \textit{magic state} with a sufficiently
low error rate \cite{Bravyi2005}.
\textit{Magic state distillation} is a family of protocols used to generate magic states with high fidelity.
While other approaches exist, such as code switching to obtain universality on FTQC\cite{Paetznick2013FaultTolerant, Anderson2014FaultTolerantConversion, Bombín2015GaugeColorCodes, Bombín2016DimentionalJump}, magic state distillation is deemed to be one of the most promising approaches \cite{Beverland2021CostOfUniversality}.
Magic state distillation is a probabilistic process, and fails with a small probability.
When it fails, the system retries distillation.
FTQC programs consume many magic states,
and magic state distillation is one of the major bottlenecks of FTQC.

To run a big quantum program such as factoring on a large FTQC system, the system needs many optimizations.
A major optimization strategy is to take advantage of parallelism to use hardware resources efficiently.
Such optimizations are often performed at compile-time; FTQC compilers reorder and schedule operations in the input program so that they can run in parallel.
However, scheduling operations that have non-deterministic durations is difficult in general.
Magic state distillation is one such operation because of its probabilistic nature.
When a magic state distillation operation fails, the system needs to stop computation
that depends on the magic state (i.e., running only the logical identity operation with error correction).
The more parallelized that computation, the bigger the cost of failure.
We can rephrase this using Amdahl's law \cite{Amdahl1967}, which states that the
performance improvement gained by parallelization is limited by the sequential portion of the computation.
If the failure case is poorly parallelized, this limits overall performance.

Given the importance of magic states in FTQC,
many studies have been conducted both on the production and consumption sides.
On the production side, the spatial and time costs of magic state distillation have been studied extensively \cite{Campbell2017Unified,Gidney2019EfficientMagicState,Litinski2019magicstate,Litinski2019GameOfSurfaceCodes,Herr2017LatticeSurgery}.
On the consumption side, many approaches have been proposed that aim for low T depth, that is, the number of magic states that cannot be performed in parallel
\cite{Amy2013MeetInTheMiddle,Selinger2013Quantum,Amy2014Polynomial,Yoshioka2022Hunting}.
They exhibit the desire to consume magic states in parallel.

In this paper, we propose introducing a pool of magic states next to each magic state factory to mitigate the negative performance impact caused by magic state distillation failures on a parallel-processing quantum computer.
To verify the magnitude of the negative impact and the usefulness of the mitigation approach, we ran two types of simulations.
One is the random quantum circuit, which is inspired by random quantum circuit sampling \cite{Bouland2018OnTheComplexity, Arute2019QuantumSupremacy}.
The other is distributed SELECT operation \cite{Babbush2018,Yoshioka2022Hunting}, which is an essential component of key quantum algorithms such as the phase estimation \cite{Kitaev1995Quantum} and the quantum singular value transformation \cite{Martyn2021GrandUnification} used in a wide variety of applications including quantum chemistry\cite{Aspuru-Guzik2005SimulatedQuantumComputation}.

Our contributions are summarized as follows:
\begin{itemize}
  \item We indicate that the impact of magic state distillation failures is amplified by parallel processing, and we quantified the impact.
        Specifically, in the random quantum circuit simulation with a distillation failure probability of 5\%, the run-time delay caused by distillation failures was comparable with increasing the  distillation time cost over 100\%.
        In the Dist-SELECT simulation, the run-time delay was comparable with increasing the distillation time cost over 50\% while the level-1 magic state distillation failure probability was approximately 5\%.
  \item We propose pooling magic states to mitigate the negative performance impact caused by magic state distillation failures on a parallel-processing quantum computer.
        Our experiments showed that pooling effectively reduced the run-time delay, with some additional spatial cost.
        Specifically, in the Dist-SELECT simulation, pooling illuminated much of the run-time delay, with an additional spatial cost of 9\%.
        We also observed that pooling reduced the number of qubits required for distillation by 45\% without increasing the time cost.
\end{itemize}

Our conclusion is that magic state distillation failures could add a large run-time delay, particularly when the circuit is highly parallelized, and system designers and/or programmers need to pay attention to the failures.
System designers can use pooling to mitigate the run-time delay.

In \autoref{sec:Background}, we define terms and concepts used in this paper.
In \autoref{sec:Motivation}, we describe our motivation and the problem that we want to solve.
In \autoref{sec:Mitigations}, we discuss how to mitigate the negative effect of magic state distillation failures.
In \autoref{sec:Performance evaluation}, we describe the simulations we ran to evaluate the mitigation approach.
In \autoref{sec:Conclusion}, we present our conclusion.

\section {Background}
\label{sec:Background}

In this section, we define terms and concepts used in this paper.
We also describe our model of quantum computers.

\subsection{Basics of fault-tolerant quantum computing}
\label{subsec:background-basics}

There are various hardware proposals for quantum computing.
In this paper, we focus on quantum computers that have nearest-neighbor connectivity such as superconducting quantum computers.
We use the surface code \cite{Kitaev2003,Bravyi1998} for the error correcting code since it is being actively developed because of its relatively high error threshold with nearest-neighbor connectivity.
With the surface code, multiple physical qubits form one logical qubit.
We call the error probability for each physical qubit the \textit{physical error probability}.

We use the time required to perform an error syndrome measurement (\textit{cycle time}) as the time unit.
For example, $3d$ cycles are required to perform a logical Hadamard operation where $d$ is the code distance of the surface code.
For the ease of analysis, we ignore the time cost of very light logical operations, such as $\ket{0}$ initialization.
Some operations, such as lattice-merge \cite{Horsman2012}, produce byproduct Pauli operators depending on the measurement results,
but in most cases, we can just track them in the
controlling classical computer and use them to reinterpret the measurement results, instead of actually performing them on the quantum computer \cite{Riesebos2017}.

\begin{figure}[t!]
\begin{minipage}[t!]{2.5cm}
 \centering
  \[
  \Qcircuit @C=.4em @R=.8em {
    \lstick{q_1} & \qw       & \qw      & \ctrl{3} & \qw \\
    \lstick{q_2} & \qw       & \ctrl{1} & \qw      & \qw \\
    \lstick{q_3} & \gate{Y}  & \targ    & \qw      & \qw \\
    \lstick{q_4} & \gate{H}  & \qw      & \targ    & \qw \\
    }
  \]
\end{minipage}
\begin{minipage}[t!]{4.5cm}
 \includegraphics[width=4.0cm]{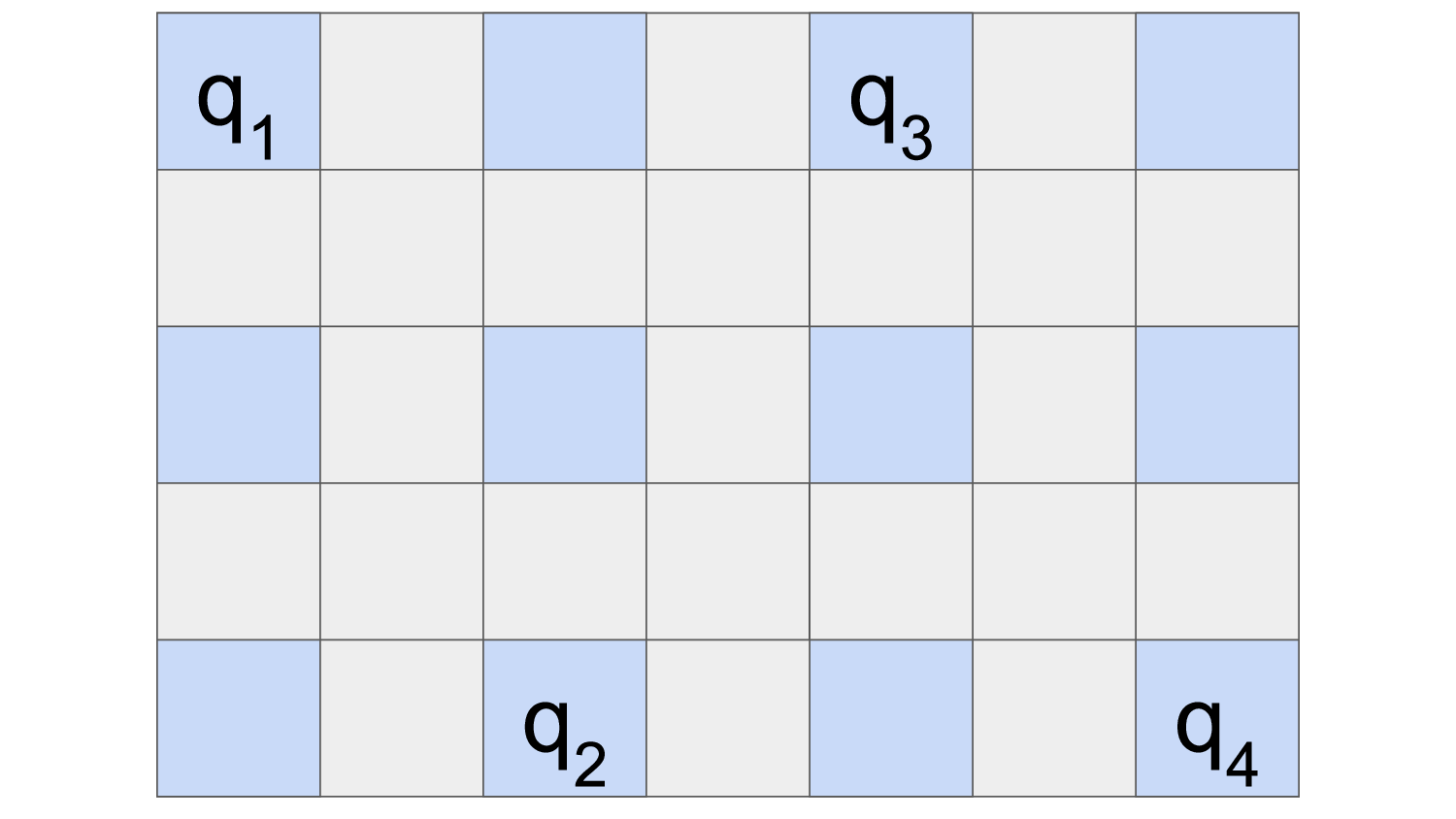}
\end{minipage}
 \caption{Example circuit (left) and placement of qubits on a qubit chip (right).
 Logical qubits can be placed on blue squares and the gray area can be used to form routes.}
 \label{fig:placement-and-routing}
\end{figure}

\autoref{fig:placement-and-routing} (left) is an abstract quantum circuit.
Our target computer can run commuting operations in parallel.
A trivial example in the figure is a pair of the $Y$ operation on $q_3$ and the $H$ operation on $q_4$.
They commute and can run in parallel.
The abstract circuit assumes all-to-all connectivity but we need to find a route between the qubits to perform a two-qubit gate on the target computer, which is a gap that compilers need to fill.
For example, in \autoref{fig:placement-and-routing}, $q_2$ and $q_3$ are placed distantly, and hence we need to create and consume a chain of Bell pairs between them to perform a two-qubit operation, $CX$, on them.
Many possible routes exist between two qubits.
The process of choosing a route is called \textit{routing}.
This is performed with \textit{placement}, the process to associate each qubit in the abstract circuit with an actual logical qubit in the computer.
Routing and placement are major functionalities of compilers targeting this type of computer \cite{Beverland2022EdgeDisjointPath}.
They may prevent some operations from running in parallel.
For example, the placement in \autoref{fig:placement-and-routing}
prevents $CX_{1, 4}$ and $CX_{2, 3}$ from running in parallel with any routes.

\subsection {Magic state distillation}
\label{subsec:background-magic-state-distillation}
The basic gates of FTQC consist of two types of gates: Clifford gates and non-Clifford gates.
Clifford gates are relatively cheap to implement fault-tolerantly in terms of space and time costs.
To implement a non-Clifford gate fault-tolerantly,  we need a special quantum state called a \textit{magic state} with a sufficiently low error rate\cite{Bravyi2005}.
\textit{Magic state distillation} is a family of protocols used to generate magic states by
distilling multiple low-quality magic states into fewer high-quality magic states.
Two types of undesirable outcomes are generated by a distillation protocol.
One is \textit{failure} that can be detected in the distillation protocol.
When a failure is detected, the system retries distillation, which makes the distillation process non-deterministic.
The other is \textit{error} that cannot be detected in the distillation protocol.
It is generally difficult to recover from errors, and the system needs to choose a distillation protocol that has a sufficiently low error rate.
We write the error rate as $p_{\mathrm{out}}$ and the failure rate as $p_{\mathrm{fail}}$.
Our focus in this paper is failure rather than error.
A \textit{Repeat-Until-Success} (RUS) operation \cite{Paetznick2014Repeat} is an operation which fails with a non-zero probability.
When it fails, the system retries the operation until it succeeds.
Magic state distillation is a RUS operation, as described above.

There are methods to generate magic states non-fault-tolerantly, that is, with an
error probability proportional to the physical error probability, such as state injection and
faulty T measurements.
Such low-quality magic states are the ingredients of distillation.
For example, the 15-to-1 protocol \cite{Bravyi2005} consumes 15 low-quality magic states and generates one high-quality magic state.
Let $p$ be the error rate of the low-quality magic states, and the error rate of the high-quality magic state is approximately $35p^3$.
This is called level-1 distillation.
If $35p^3$ is not sufficiently small, we can distill magic states generated by level-1 distillation to generate even higher-quality magic states.
This is called level-2 distillation.
Level-$n$ distillation for any $n$ is defined in the same manner.

The computation in a magic state distillation protocol is also protected by an error correcting code, but it is not necessarily the same as the code used outside of the protocol.
For example, when implementing a two-level distillation protocol,
it is reasonable to use a high-quality (and expensive) error code for the level-2 distillation part
and use a low-quality (and cheap) error code for the level-1 distillation part,
given that the level-1 distillation part generates relatively low-quality
magic states in any case.
Moreover, Litinski \cite{Litinski2019magicstate} used different distances
depending on orientation.
When we use the surface code with distance $d$, we use the same distance for
two spatial axes ($X$-axis and $Z$-axis) and the time axis.
Instead, they used different distances to save the spatial and time costs of distillation.
In this paper, we use such distillation protocols.

\begin{figure}[t!]
 \centering
 \includegraphics[width=8cm]{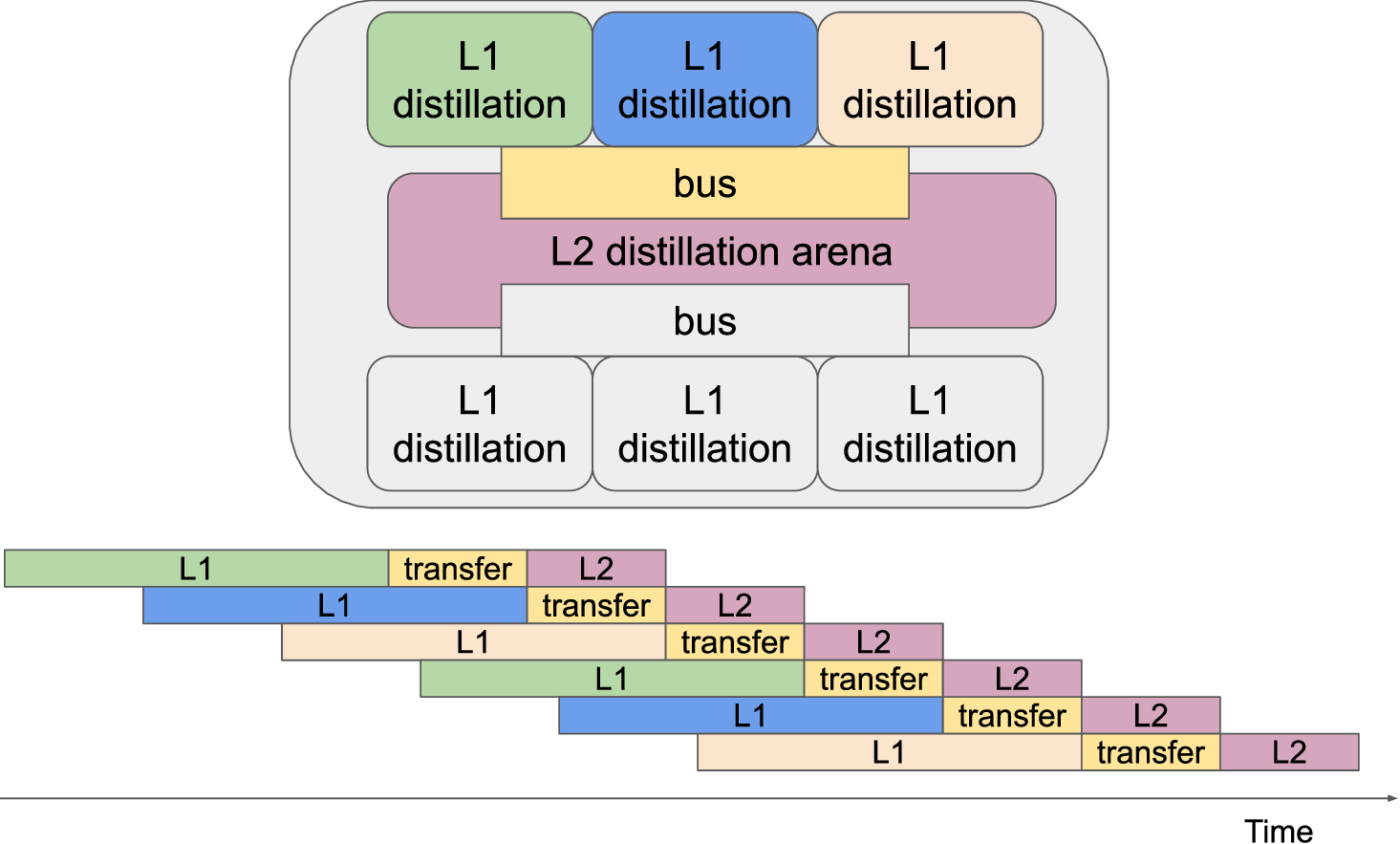}
 \caption{Block diagram for a two-level 15-to-1 distillation protocol (top) and one of its pipelines (bottom). Each colored box in the pipeline is a process hosted by the same-color sub-circuit in the block diagram.
 ``L1'' represents L1 distillation, ``transfer'' represents the L1 magic state transfer process, and ``L2'' represents the part of L2 distillation using the L1 magic state.}
 \label{fig:magic-state-distillation-pipeline}
\end{figure}

Two-level distillation protocols described in Ref. \cite{Litinski2019magicstate} extensively use pipelining to reduce the distillation time cost.
For example, a two-level 15-to-1 protocol consists of 1) $n_{L1}$ level-1 distillation blocks, 2) one level-2 distillation arena, and 3) two buses between the level-1 distillation blocks and the level-2 distillation arena for some even integer $n_{L1}$.
The level-2 distillation arena can consume two level-1 magic states simultaneously; hence, we have two pipelines, each of which contains $n_{L1}/2$ level-1 distillation blocks.
\autoref{fig:magic-state-distillation-pipeline} illustrates such a pipeline with $n_{L1} = 6$.

\subsection{Parallel execution}
\label{subsec:background-parallel-execution}

Running a large program on a parallel-processing computer efficiently is difficult.
A simple solution is to run one operation at a time, but this is not efficient.
Another simple solution is scheduling operations somehow, assigning a starting clock for each operation and running them with synchronizing all the physical qubits at each clock.
This would not scale even without RUS operations, because the synchronization cost is very large on large computers such as one having 20 million physical qubits\cite{Gidney2021howtofactorbit}.
With RUS operations the scenario becomes worse.
When a RUS operation fails, the system needs to stop all operations except for the retrying RUS operation, which adds a significant time cost to the failure.

Zhang~\textit{et al.}~\cite{Zhang2021} extracted a dependency graph from an input program and performed operations according to the dependency graph.
This is sufficiently flexible to handle RUS operations.
We use this execution model with two notes.

One is regarding routing.
As described above, our target computer requires qubit placement and routing at compile-time.
They introduce additional dependencies among operations and make the dependency graph much bigger.
For example, if we have a $CX$ operation on two qubits 100 logical qubits away,
we require a route between them, which means we have no less than 100 additional nodes in the dependency graph.

The other note is regarding retrying.
Retrying a RUS operation could be complicated because, elsewhere in the model, time flows monotonically, but when a RUS operation fails, time rolls back for the qubits involved in the operation (including routing qubits).
The scenario becomes more complicated when we consider nested failures,
which is possible in two-level magic state distillation.
We avoid this problem by exempting magic state distillation for this model.
A magic state factory continuously runs distillation without looking at the dependency graph.
At the scheduling time, the scheduler distinguishes each distillation and allocates a distillation to a magic state request.
By contrast, at run-time, a consumer does not distinguish each distillation,
and uses the earliest available magic state generated by the assigned magic
state factory while conforming to the explicit and implicit dependencies.
This also solves the nesting problem.
This is justifiable given that magic state distillation has no input, and we assume
magic state factories are placed statically and their internal routing does not
conflict with external routing.
The importance of magic state distillation is another justification for this special treatment.

\subsection{SELECT and Dist-SELECT circuits}
\label{subsec:background-select-and-dist-select}
\begin{figure}[h!]
  \centering
  \[
  \Qcircuit @C=.4em @R=.8em {
    \lstick{ctrl} & \ctrl{1} & \qw & \qw & \qw & \qw & \qw & \qw                  & \ctrl{2}  & \qw & \qw & \qw & \qw & \qw & \qw & \ctrl{1}   \\
    \lstick{l_0}  & \ctrlo{1} & \qw & \qw & \qw & \qw & \qw & \qw                 & \qw       & \qw & \qw & \qw & \qw & \qw & \qw & \ctrl{1} \\
    \lstick{a_0}  & \qwx & \ctrl{1} & \qw & \ctrl{2} & \qw  & \ctrl{1} & \qw      & \targ     & \ctrl{1} & \qw & \ctrl{2} & \qw  & \ctrl{1} & \qw & \qw \\
    \lstick{l_1}  & \qw & \ctrlo{1} & \qw & \qw & \qw & \ctrl{1} & \qw            & \qw       & \ctrlo{1} & \qw & \qw & \qw & \ctrl{1} & \qw & \qw \\
    \lstick{a_1}  & & \qwx & \ctrl{1} & \targ & \ctrl{1} & \qw &                  &           & \qwx & \ctrl{1} & \targ & \ctrl{1} & \qw & & \\
    \lstick{\ket{\psi}} & \qw & \qw & \gate{H_0} & \qw & \gate{H_1} & \qw & \qw   & \qw       & \qw & \gate{H_2} & \qw & \gate{H_3} & \qw & \qw &\qw \\
    }
  \]
  \caption{SELECT circuit with $L$ = 4.}
  \label{fig:optimized-select-circuit}
 \end{figure}
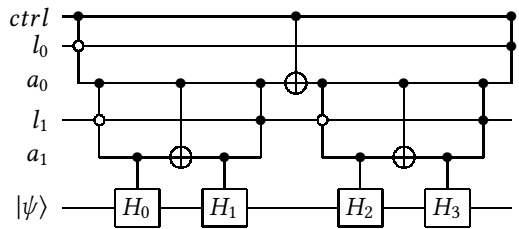

In this subsection we introduce SELECT and Dist-SELECT circuits because we simulate the circuits in \autoref{sec:Performance evaluation} (The random quantum circuit sampling, which is also used in \autoref{sec:Performance evaluation}, will be directly discussed there because it is simple.)
The SELECT oracle \cite{Babbush2018} is an important oracle used in the phase estimation algorithm \cite{Kitaev1995Quantum} and
the quantum singular value transformation algorithm \cite{Martyn2021GrandUnification},
both of which have a wide variety of applications, including quantum chemistry\cite{Aspuru-Guzik2005SimulatedQuantumComputation}:

\[
  \mathrm{SELECT} \coloneqq \sum_{\ell=0}^{L - 1} \op{\ell}{\ell} \otimes H_\ell,
\]
where each $\ell$ is an integer expressed with $\log L$ qubits (\textit{index qubits})
and each $H_\ell$ is a Pauli operator that operates on $N$ qubits (\textit{Pauli target qubits}).
For simplicity, we assume that $L$ is a power of two.
We call $L$ the length of the SELECT oracle.
We can naively implement the oracle with $\mathcal{O}(L \log L)$ Toffoli gates and $\log L$ ancilla qubits.
Babbush~\textit{et al.}~\cite{Babbush2018} proposed an implementation with $\mathcal{O}(L)$ Toffoli gates (\autoref{fig:optimized-select-circuit}).

Yoshioka~\textit{et al.}~\cite{Yoshioka2022Hunting} proposed a distributed SELECT circuit (Dist-SELECT)
which shortens the circuit execution time by parallelizing the execution.
Let $M$ be a power two less than $L$.
Dist-SELECT with parallelism $M$ consists of $M$ sub-circuits, each of which has $\log \frac{L}{M}$ index qubits and $\log \frac{L}{M}$ ancilla qubits.
All the sub-circuits share the Pauli target qubits and each sub-circuit runs a SELECT operation with length $\frac{L}{M}$.
This increases the number of qubits because of the additional index and ancilla qubits, but as long
as $M$ is sufficiently smaller than $N$, the additional spatial cost is relatively small.
At the expense of the additional qubits, each SELECT operation has a shorter length, which may lead to a shorter execution time.

Because all the sub-circuits share the Pauli target qubits, actual parallelism depends on how many controlled Pauli operators we can run in parallel on the Pauli target qubits.
This depends on 1) each Pauli operator $H_\ell$ determined by the problem itself, and 2) the routing efficiency determined by the quantum computer architecture and the compiler.

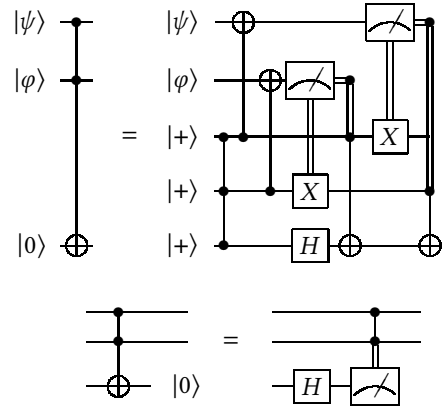
\begin{figure}[h!]
\centering
\[
\Qcircuit @C=.2em @R=.8em {
  \lstick {\ket{\psi}}    & \ctrl{1} & \qw & \push{\rule{1.0em}{0em}} &   & \push{\rule{1.0em}{0em}\ket\psi\rule{.4em}{0em}}    & \qw        & \targ     & \qw       & \qw             & \qw       & \meter \cwx[2] & \cctrl{3} & \\
  \lstick {\ket{\varphi}} & \ctrl{3} & \qw & \push{\rule{1.0em}{0em}} &   & \push{\rule{1.0em}{0em}\ket\varphi\rule{.4em}{0em}} & \qw        & \qw       & \targ     & \meter \cwx[2]  & \cctrl{1} &                &           & \\
                          &          &     & \push{\rule{1.0em}{0em}} & = & \push{\rule{1.0em}{0em}\ket+\rule{.4em}{0em}}       & \ctrl{1}   & \ctrl{-2} & \qw       & \qw             & \ctrl{2}  & \gate{X}       & \qw       & \\
                          &          &     & \push{\rule{1.0em}{0em}} &   & \push{\rule{1.0em}{0em}\ket+\rule{.4em}{0em}}       & \ctrl{1}   & \qw       & \ctrl{-2} & \gate{X}        & \qw       & \qw            & \ctrl{1}  & \\
  \lstick{\ket{0}}        & \targ    & \qw & \push{\rule{1.0em}{0em}} &   & \push{\rule{1.0em}{0em}\ket+\rule{.4em}{0em}}       & \ctrl{-1}  & \qw       & \qw       & \gate{H}        & \targ     & \qw            & \targ     & \\
  }
\]
\\
\[
\Qcircuit @C=.8em @R=.8em {
  \lstick{} & \ctrl{1} & \qw & \qw            &          & \push{\rule{.5em}{0em}} & \qw      & \ctrl{1}    & \qw \\
  \lstick{} & \ctrl{1} & \qw & \qw            & \push{=} & \push{\rule{.5em}{0em}} & \qw      & \ctrl{-1}   & \qw \\
  \lstick{} & \targ    & \qw & \push{\ket{0}} &          & \push{\rule{.5em}{0em}} & \gate{H} & \meter \cwx &     \\
  }
\]

\caption{Toffoli gates used in the SELECT circuit.}
\label{fig:toffoli-gates}
\end{figure}

In the optimized SELECT circuit (\autoref{fig:optimized-select-circuit}), we use two types of Toffoli gates:
a Toffoli gate with a $\ket0$ input target qubit and its inverse.
\autoref{fig:toffoli-gates} shows their implementations.
The former is implemented with gate teleportation and a CCZ resource state $\ket{CCZ} \coloneqq CCZ \ket{+}^{\otimes 3}$.
The latter is implemented without magic states.

\section {Motivation}
\label{sec:Motivation}

In this section, we show that parallel processing amplifies the impact of RUS failures, and they must be mitigated for high-performance FTQCs.
In the following, we describe our motivation, formulate the problem we want to solve, and list related studies.
First, we analytically evaluate the impact of failures in sequential and parallel executions of RUS operations.
Let $U$ be a RUS operation with a failure probability $p$.
The expected execution time required for $n$ sequential $U$ operations is
$n \sum_{i = 0}^{\infty}{p^i} = \frac{n}{1 - p}$.
The expected execution time required for $n$ parallel $U$ operations, say $\{a_n\}$, satisfies $a_0 = 0$ and  $a_n = 1 + \sum_{i = 0}^{n} \binom{n}{i} (1 - p)^{n - i} p^i a_i$.

\begin{table}[h!]
  \centering
  \caption{Expected execution time with $p$ = 0.01.}
  \label{tab:sequential-parallel-rus-execution-time}
  \begin{tabular}{|r|r|r|}
    \hline
    \textbf{$n$} & \textbf{sequential (increase)} & \textbf{parallel (increase)} \\
    \hline
    \hline
      1 & 1.0101 (1.01\%) & 1.0101 (1.01\%) \\
    \hline
      10 & 10.101 (1.01\%) & 1.0967 (9.67\%) \\
    \hline
      100 & 101.01 (1.01\%) & 1.6440 (64.4\%) \\
    \hline
  \end{tabular}
\end{table}

\autoref{tab:sequential-parallel-rus-execution-time} shows the expected execution time for $n$ sequential and parallel $U$ operations with $p = 0.01$ and the relative increase over the execution time with $p = 0$.
Failures have little impact on sequential executions, but they have a much larger impact on parallel executions.
This is because, by exploiting parallelism, parallel execution is more efficient than sequential execution, whereas the failure case is poorly parallelized.
An issue that arises is whether the 64\% increase in the parallel case matters given that the parallel case is much more efficient than the sequential case.
We would like to emphasize that failures could limit the benefit of parallelization, even when the failure probability is small: if the failure probability is 0, 100-parallelization would bring a 100x speed up, but with 1\% failure probability, it \textit{only} brings 61x speedup.
In this regard, our baseline is the parallel case with $p = 0$ rather than the sequential case.

\begin{figure}[t!]
 \centering
 \includegraphics[width=8.5cm]{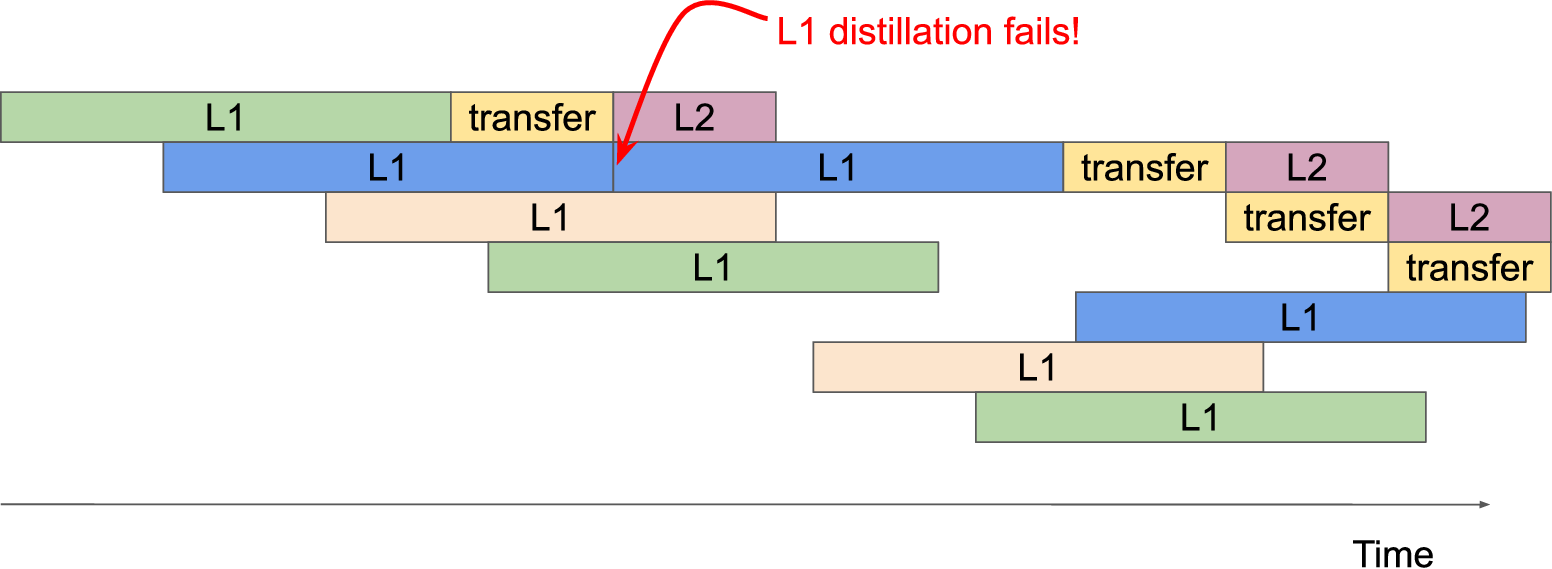}
 \caption{Pipeline with an L1 distillation failure.}
 \label{fig:magic-state-distillation-broken-pipeline}
 \end{figure}

Pipelining is an effective technique for using hardware resources efficiently by capitalizing on parallelism.
As described in \autoref{subsec:background-magic-state-distillation}, the magic state distillation protocols we use leverage pipelining to minimize the distillation time cost.
Consider, for example, a two-level 15-to-1 distillation circuit with $n_{L1} = 6$.
\autoref{fig:magic-state-distillation-pipeline} shows one of two pipelines in such a circuit.
The impact of level-1 distillation failures could be amplified by pipelining, as depicted in \autoref{fig:magic-state-distillation-broken-pipeline}.
The longer the pipeline, the greater the cost of failures.

In this section, we have presented two examples that contain RUS operations.
In both cases, parallelization plays a key role in shortening the execution time, but it also amplifies the negative effect of RUS failures.
We believe that this scenario is generic when a quantum program is highly parallelized, and this problem requires attention given that FTQC programs consume many magic states during computation.

The spatial and time costs of magic state distillation have been studied extensively\cite{Campbell2017Unified,Gidney2019EfficientMagicState,Litinski2019magicstate,Litinski2019GameOfSurfaceCodes,Herr2017LatticeSurgery}.
On the magic state consumption side, many approaches have been proposed that aim for low T depth, that is, the number of magic states that cannot be performed in parallel
\cite{Amy2013MeetInTheMiddle,Selinger2013Quantum,Amy2014Polynomial,Yoshioka2022Hunting}.
They exhibit the desire to consume magic states in parallel.

Ref.\cite{Bombín2024FaultTolerant} proposed placing a pool of magic states in each distillation factory to mitigate magic state injection failures.
Its main target is photonic FBQC architectures which are different from our target architectures.
Ref.\cite{Beverland2022Assessing} proposed placing excessive L1 distillation factories in each L2 distillation factory to mitigate L1 magic state distillation failures.
In \autoref{sec:Mitigations}, we present our solution, which is placing an external pool of magic states to each distillation factory.

\begin{figure}[t!]
 \centering
 \includegraphics[width=8.0cm]{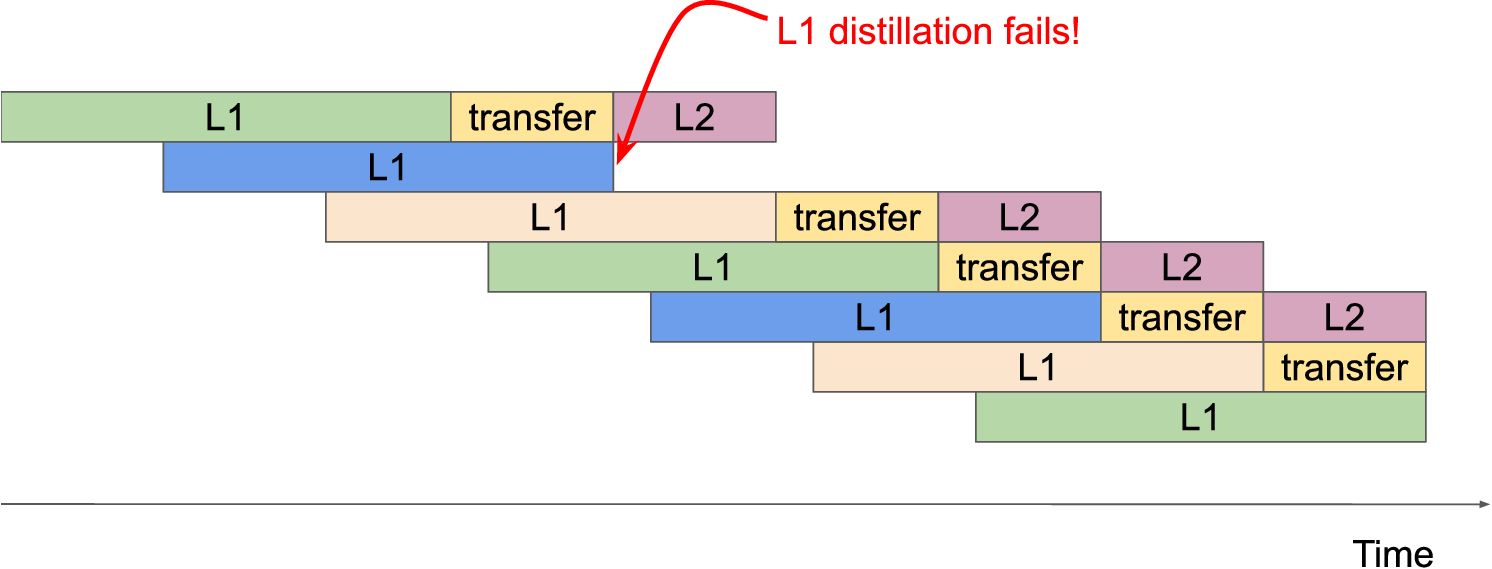}
 \caption{Pipeline with an L1 distillation failure, mitigated.}
 \label{fig:magic-state-distillation-racing-pipeline}
\end{figure}

For two-level distillation protocols, we can mitigate the effect of L1 distillation failures relatively easily.
When an L1 distillation fails, instead of waiting for the original L1 distillation factory, we can choose the best L1 distillation factory in the pipeline at run-time, as shown in \autoref{fig:magic-state-distillation-racing-pipeline}.
This mitigation approach is possible because the L1 distillation factories generate identical (albeit noisy) magic states, and the concrete circuit layout allows the run-time factory switching.
Note that this approach does not work in general scenarios.
As discussed in \autoref{subsec:background-parallel-execution}, routing and the computational dependency graph are intertwined, and the approach incurs a prohibitively high classical computational cost in large circuits.
The purpose of this study is to find a mitigation approach applicable in more general scenarios.

To fully test our statement and the solutions described in \autoref{sec:Mitigations}, we require an FTQC program, FTQC, and FTQC compiler.
Many FTQC algorithms exist; however, no FTQC nor FTQC compilers are available at the present time.
Hence, we use the simulations explained in \autoref{sec:Performance evaluation}.

\section{Dealing with distillation failures}
\label{sec:Mitigations}

In this section, we discuss an approach to mitigate the negative effect of magic state distillation failures.
We have two metrics: spatial cost (number of qubits) and time cost (number of cycles).
We call the time cost known at the scheduling phase the \textit{scheduled time cost}.
Magic state distillation failures increase the time cost at run-time; hence, we call it the \textit{run-time delay}.
The goal of the approach is to reduce the total time cost while minimizing the additional spatial cost.

To assess the impact of magic state distillation failures, we introduce additional terms.
Let us assume we perform a computational task with a distillation protocol whose distillation time cost is $D$, the scheduled time cost is $S$, and the average run-time delay is $R$.
The term \textit{relative run-time delay} refers to $\frac{R}{S}$, which is the run-time delay relative to the scheduled time cost.
Although this quantifies the amount of the impact of distillation failures, we often want to know the amount of the impact in relation to the distillation time cost.
Let us also assume we perform the same task with another distillation protocol whose distillation time cost is $E$ and the scheduled time cost equals $S + R$.
Let the \textit{effective distillation time cost} of the original distillation protocol be $E$\footnote{The scheduled time cost tends to increase as the distillation time cost increases, but it is not a completely increasing function. Therefore, multiple $E$ values could exist. In such a case, we select the largest value.}.
This can be understood as the distillation time cost with the run-time delay incorporated: the original distillation protocol and a hypothetical distillation protocol with time cost $E$ and no failure probability lead to the same total time cost.
We further define the \textit{relative distillation time cost increase} as $\frac{E - D}{D}$.
For example, let us suppose the original distillation time cost is 100, the scheduled time cost for the original distillation time cost is 1200, the run-time delay is 300, and the scheduled time cost for the distillation time cost 200 is 1500.
Then the relative run-time delay is 25\% and the relative distillation time cost increase is 100\%.
In this case the run-time delay effectively doubles the distillation time cost.

To hide the latency caused by distillation failures, we introduce a pool of magic states, which is a first-in-first-out queue, to each magic state factory.
We expect the pool to work as a buffer between the factory and the consumers of magic states.
Each factory continuously distills magic states and the generated magic states are stored in the pool.
On the consumption side, at the arrival of a magic state request, the pooled magic state is transferred from the pool to the consumer, if possible: otherwise, the request needs to wait for the factory.
Note that this behavior is dynamic, similar to magic state distillation.
This is aligned with the execution model discussion in \autoref{subsec:background-parallel-execution}.
The additional spatial cost is the number of qubits required for the pool.
In a pool, using a smaller code distance than that for usual data qubits may be acceptable in terms of the error probability because magic states remain there for a relatively short period of time.

\begin{figure}[h!]
  \centering
  \[
  \Qcircuit @C=.4em @R=.8em {
                        & \control                       & \cctrl{2} \cw & \\
    \lstick{\ket{\psi}} & \multigate{1}{M_{ZZ}} \cwx[-1] & \qw           & \gate{M_{X}} \cwx[1] & \control                       & \cctrl{2} & \\
    \lstick{\ket{+}}    & \ghost{M_{ZZ}}                 & \gate{Z}      & \gate{Z}             & \multigate{1}{M_{ZZ}} \cwx[-1] & \qw       & \gate{M_{X}} \cwx[1]  \\
    \lstick{\ket{+}}    & \qw                            & \qw           & \qw                  & \ghost{M_{ZZ}}                 & \gate{Z}  & \gate{Z} & \qw & \rstick{\ket{\psi}} \\
    }
  \]
  \caption{Transferring a quantum state. $M_{ZZ}$ denotes the $ZZ$ measurement and $M_{X}$ denotes the $X$ measurement.}
  \label{fig:qubit-transfer}
\end{figure}
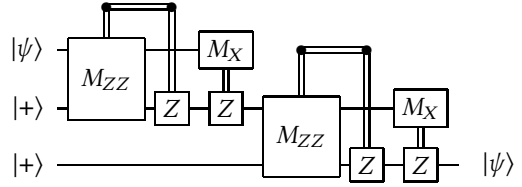

\begin{figure}[!t]
  \centering
  \includegraphics[width=6.5cm]{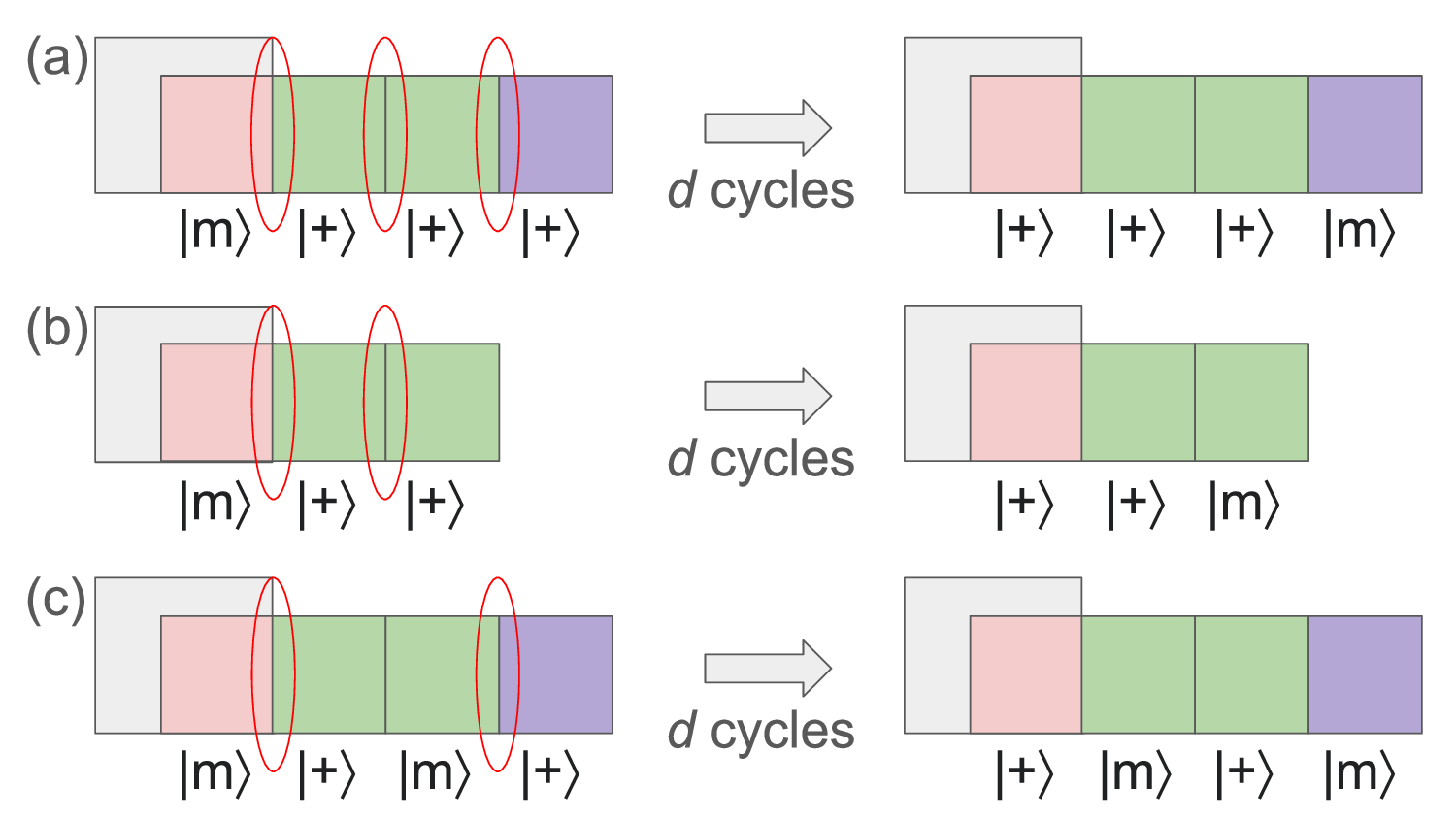}
  \caption{Transferring quantum states across a pool with two entries.
  Different scenarios presented in rows demonstrating the status of qubits before (left) and after (right) the transfer.
  Pink, green, and purple squares are qubits in magic state factories, pools, and the consumers of magic states, respectively.
  Each red ellipse is a lattice surgery operation and $\ket{m}$ is a magic state.
  The routing qubits between the pool and the consumer are omitted.
  (a) The pool is empty and the magic state is transferred directly to the consumer.
  (b) There is no magic state request and the magic state is stored in the pool.
  (c) A pooled magic state is transferred to the consumer while another magic state is stored in the pool.}
  \label{fig:pooling parallel state transfer}
\end{figure}

\begin{figure}[!t]
  \centering
  \includegraphics[width=8.2cm]{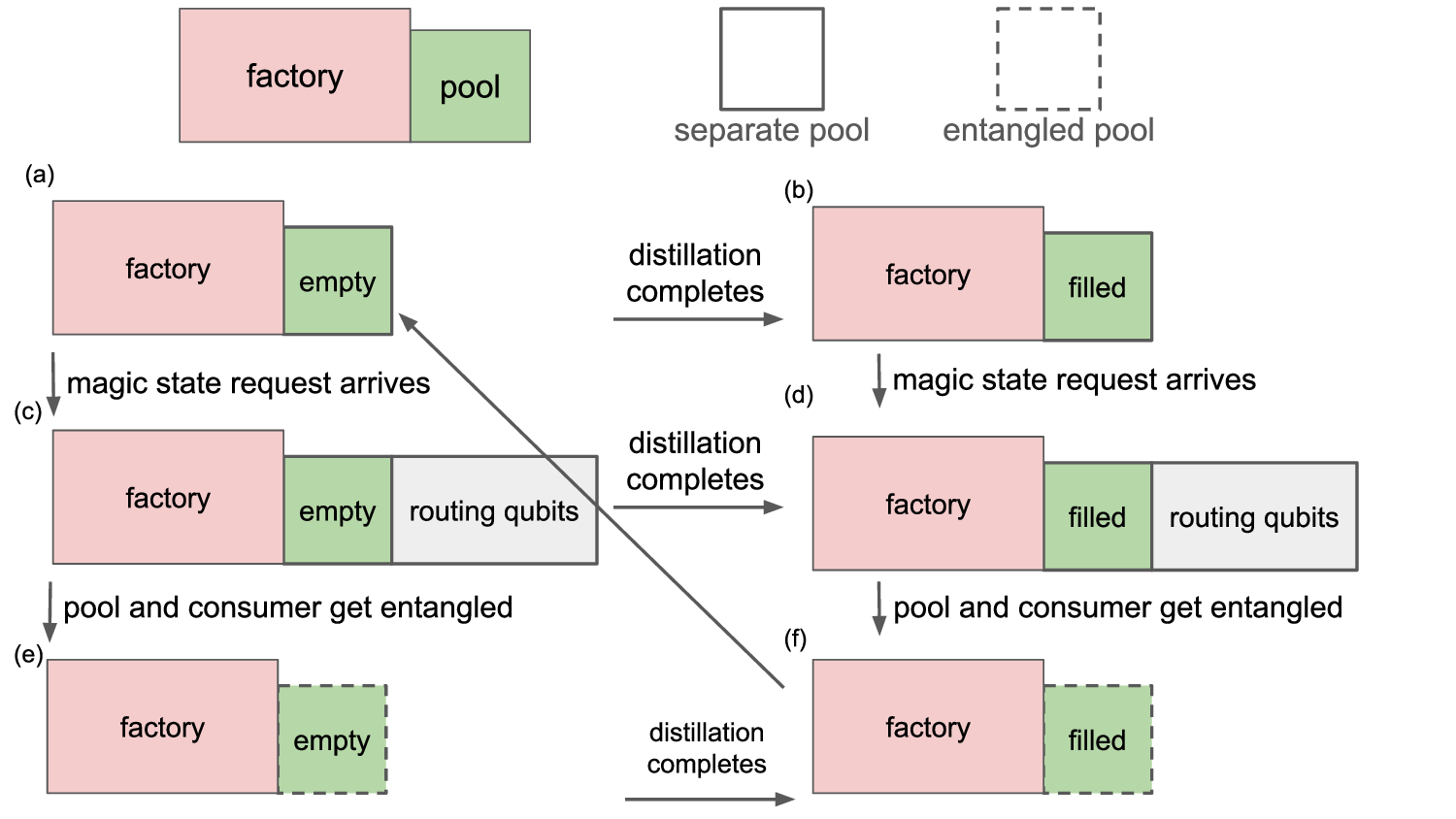}
  \caption{Run-time behavior of pooling illustrated as a state diagram.
   A pool is entangled with the consumer if it has dashed boundaries.
  (a)--(f) represent a factory with
  (a) an empty pool,
  (b) a filled pool,
  (c) an empty pool being entangled with the consumer (through the routing qubits),
  (d) a filled pool being entangled with the consumer (through the routing qubits),
  (e) an empty pool entangled with the consumer, waiting for a magic state, and
  (f) a filled pool entangled with the consumer.
  At (f), the consumer can use the magic state and the state immediately transitions to (a).}
  \label{fig:pooling runtime behavior}
\end{figure}

We transfer a quantum state across qubits with quantum teleportation, i.e., entangling the qubits and measuring the source qubit.
Entangling adjacent qubits is performed by lattice surgery \cite{Horsman2012} and hence entangling two qubits can be naively performed in $(l + 1)d$ cycles where $l$ is the length of the route between the two qubits and $d$ is the code distance, as depicted in \autoref{fig:qubit-transfer}.
Note that $Z$ gates in \autoref{fig:qubit-transfer} have no time cost, as discussed in \autoref{sec:Background}.
When the route is linearly shaped, though, the $(l + 1)$ lattice surgery operations can be carried out simultaneously and the entanglement can be established in $d$ cycles.
Hence, unlike pools on classical computers, this pool does not add latency.
The benefit of parallelization grows as the length of the queue grows.
\autoref{fig:pooling parallel state transfer} shows typical transfer scenarios and \autoref{fig:pooling runtime behavior} is a state transition diagram of single entry pooling.

\begin{figure}[!t]
  \centering
  \includegraphics[width=3.6cm]{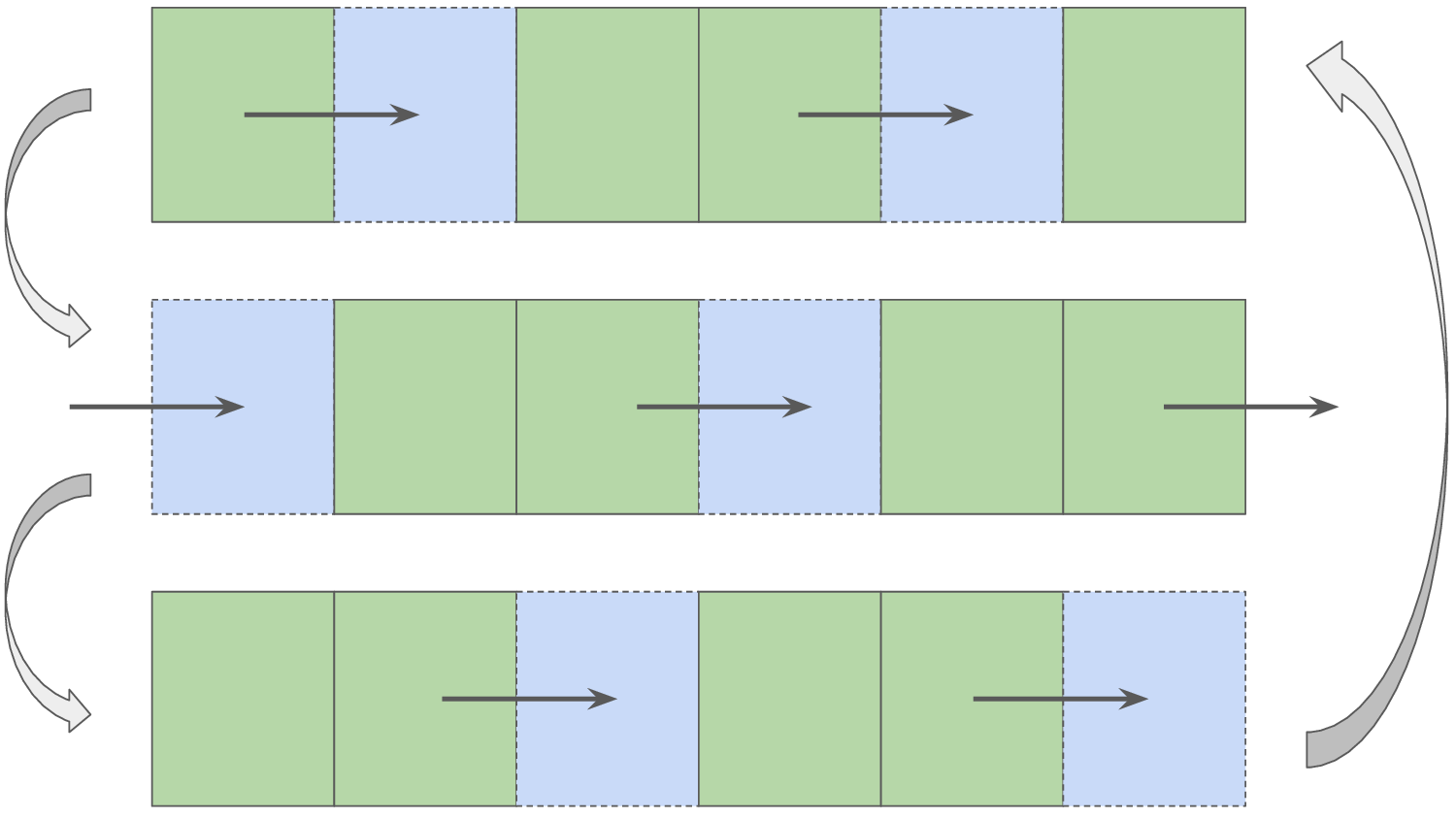}
  \caption{Pool with six entries connected to a magic state distillation factory that generates a magic state for every $3d$ cycles.
  A blue square represents an empty qubit and a green square represents a filled qubit.
  Each row represents the pool in a different time slice. 
  A state is enqueued to the left end and dequeued from the right end.}
  \label{fig:pooling with multiple queues}
\end{figure}

This discussion requires the presence of a route between the endpoint qubits, which means that the intermediate qubits do not store magic states in our context.
As a consequence, transferring quantum states in a denser pool would be more expensive.
Assuming that we distill and consume a magic state every $nd$ cycles, a pool with $nm$ entries can retain $\frac{(n - 1)m}{n}$ states on average, as depicted in \autoref{fig:pooling with multiple queues}.
The sliding scheme discussed in Ref.\cite{McEwen2023RelaxingHardware} may expedite the transferring process when the pool is densely occupied.

The speeds of magic state distillation and consumption need to be closely correlated to achieve optimal performance.
Ideally, the distillation speed should slightly outpace the consumption speed.
If consumption is faster than distillation, it will deplete the pool and the consumer of magic states will be blocked, thereby extending the execution time.
If the distillation speed is exactly equivalent to the consumption speed, an empty pool will never be replenished.
Finally, a too rapid distillation speed leads to discarded magic states.

\section{Performance evaluation}
\label{sec:Performance evaluation}

We conducted two types of simulations.
In the first, we simulated the random quantum circuit to evaluate the impact of magic state distillation failures with various failure probability settings.
The second was more practical; we simulated the distributed SELECT circuit described in \autoref{subsec:background-select-and-dist-select}.
Along with evaluating the impact of distillation failures, the primary objective of this simulation was to assess the effectiveness of the mitigation approach discussed in \autoref{sec:Mitigations}.

\subsection{Random quantum circuit simulation}

Random quantum circuit sampling is sampling from the probability distributions of randomly generated quantum circuits.
Thanks to the average-hardness of the task on classical computers \cite{Bouland2018OnTheComplexity} and the relative easiness of the experimental realization, it is used to test Quantum Supremacy, for example, Ref.\cite{Arute2019QuantumSupremacy}.
We ran a simulation of a randomly generated circuit inspired by random quantum circuit sampling.
In the simulation, we apply ``layers" to a plane of logical qubits.
Each layer consists of single qubit operators and two-qubit operators (\textit{couplings}).
\autoref{tab:random circuit simulation parameters} shows the parameters.

\begin{table}[h!]
\centering
\caption{Random quantum circuit simulation parameters}
\label{tab:random circuit simulation parameters}
\begin{tabular}{|l|l|}
  \hline
  \textbf{symbol} & \textbf{description} \\
  \hline
  \hline
  $d$ & The code distance of the surface code. \\
  \hline
  $W$ & The width of the qubit plane. \\
  \hline
  $H$ & The height of the qubit plane. \\
  \hline
  $L$ & The number of layers to be applied. \\
  \hline
  $D$ & The coupling distance. \\
  \hline
  $p_{\mathrm{fail}}$ & The failure probability of magic state distillation. \\
  \hline
\end{tabular}
\end{table}

\begin{figure}[!t]
  \centering
  \includegraphics[width=2.4cm]{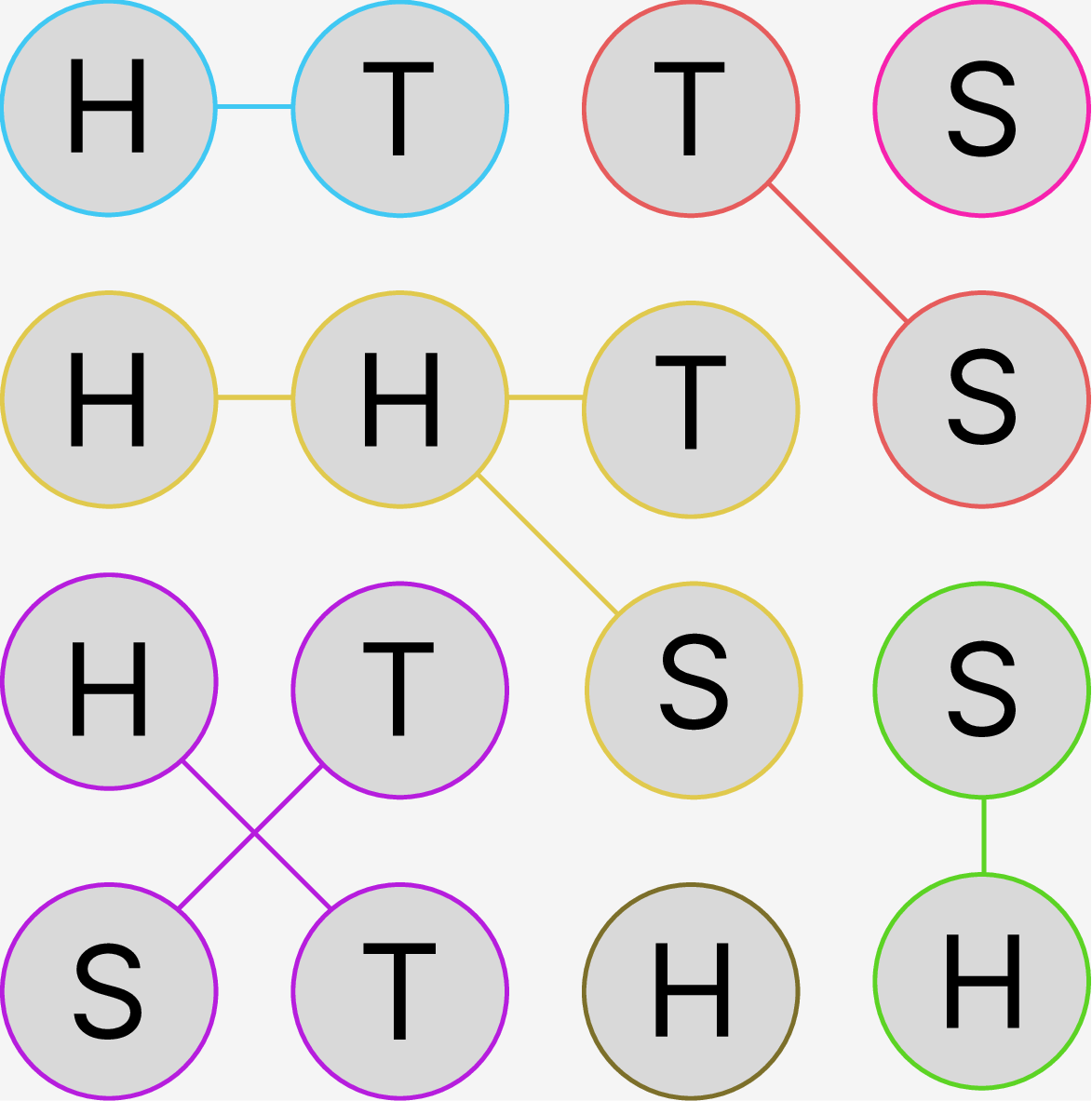}
  \caption{Layer of a random quantum circuit with $W = H = 4$, and $D = 2$.
  Each circle represents a qubit and the character in it represents a single qubit operator.
  Each line that connects two circles represents a coupling operator.
  Couplings that belong to the same coupling group have the same color, as do the qubits involved in such couplings.}
  \label{fig:random circuit layer}
\end{figure}

For each layer, we apply a randomly chosen single qubit operator ($S$, $H$, or $T$) for each qubit and then apply as many couplings as possible, with the following constraints:
\begin{itemize}
    \item Each qubit is involved in at most one coupling.
    \item Each coupling's L1 distance is less than or equal to the coupling distance parameter.
\end{itemize}
\autoref{fig:random circuit layer} shows a layer of a random quantum circuit.

\begin{table}[h!]
\centering
\caption{Time cost of operators}
\label{tab:random circuit operator time cost}
\begin{tabular}{|l|l|l|l|l|}
  \hline
  \textbf{operator} & H & S & T & coupling \\
  \hline
  \textbf{cost} & $3d$ & $2d$ & $3dNB(1, 1 - p_{\mathrm{fail}}) + 4d$ & $2d$ \\
  \hline
\end{tabular}
\end{table}

\autoref{tab:random circuit operator time cost} shows the time cost of operators in the simulation.
Using the lattice surgery, we can implement $H$ and $S$ gates with $3d$ and $2d$ cycles, respectively.
The time cost of a $T$ gate depends on the distillation protocol, but in our simulation, we simply assume distillation is a Bernoulli trial with the success probability $1 - p_{\mathrm{fail}}$ and the time cost of each trial is $3d$.
$NB(r, p)$ in the table stands for the negative binomial distribution and $3dNB(1, 1 - p_{\mathrm{fail}}) + 3d$ models the time cost of a distillation.
We also require $d$ additional cycles for the gate teleportation.
Note that we do not count the time cost of an $S$ fix-up required with the probability of 1/2, because it can be performed with zero time cost and additional ancilla qubits.
We assume that a coupling operator is a simple two-qubit operator, such as $CX$.
We can run such an operator with $2d$ cycles, even when the endpoints are distant, assuming a route between the endpoints exists.

Our simulation implicitly manages routing and distillation, adhering to the principles laid out in \autoref{subsec:background-basics} and \autoref{subsec:background-parallel-execution}.
To account for the complexity of routing, we introduce the concept of a \textit{coupling group}, which is a collection of couplings.
Two couplings belong to the same coupling group when they intersect with each other as line segments.
In \autoref{fig:random circuit layer}, couplings that belong to the same coupling group have the same color.
\autoref{alg:random-circuite-execute} is the simulation algorithm.
As in the algorithm, single qubit operators can run in parallel, whereas a coupling is bottle-necked by the slowest qubit involved in a coupling that belong to the same coupling group.
$elapsed[q]$ in the algorithm represents the elapsed cycles on qubit $q$.
The circuit's elapsed cycles are represented by the maximum elapsed value across all qubits.

\begin{algorithm}[t!]
\caption{Random circuit simulation}
\label{alg:random-circuite-execute}
\begin{algorithmic}[1]
  \FOR {$layer$ in all the layers}
    \FOR {$(q, gate)$ in single qubit gates at $layer$}
      \STATE {$elapsed[q] \leftarrow elapsed[q] + $ the time cost of $gate$}
    \ENDFOR
    \FOR {$group$ in coupling groups at $layer$}
      \STATE {$m \leftarrow 0$}
      \FOR {$(q_1, q_2)$ in $group$}
        \STATE {$m \leftarrow {\rm max}(m, elapsed[q_1], elapsed[q_2])$}
      \ENDFOR
      \FOR {$(q_1, q_2)$ in $group$}
        \STATE {$elapsed[q_1] \leftarrow m + 2d$}
        \STATE {$elapsed[q_2] \leftarrow m + 2d$}
      \ENDFOR
    \ENDFOR
  \ENDFOR
\end{algorithmic}
\end{algorithm}

\begin{figure}[t!]
  \includegraphics[width=7.0cm]{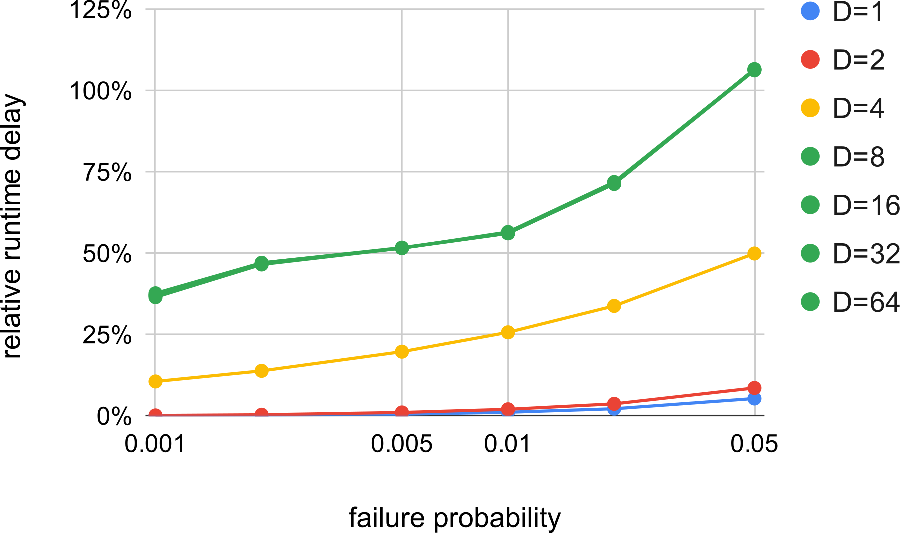}
  \caption{Relative run-time delay for each setting.
  $D \ge 8$ lines are almost identical in the graph.}
  \label{fig:random circuit relative runtime delay}
\end{figure}
\begin{figure}[t!]
  \includegraphics[width=7.0cm]{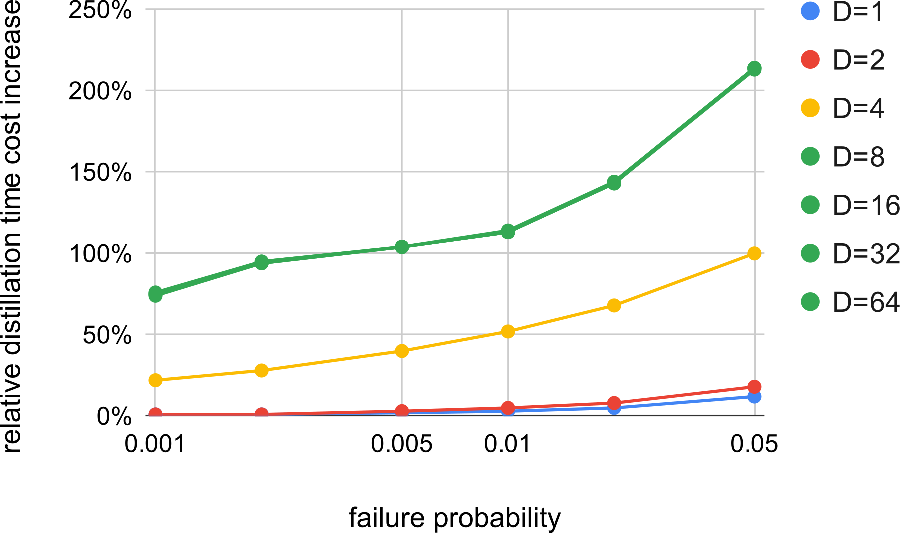}
  \caption{Relative distillation time cost increase for each setting. $D \ge 8$ lines are almost identical in the graph.}
  \label{fig:random circuit relative distillation time cost increase}
\end{figure}

We set $W = H = 64$ and $L = 100$\footnote{Because every time cost is proportional to $d$, the relative run-time delay and relative distillation time cost increase are independent of $d$.}.
\autoref{fig:random circuit relative runtime delay} and \autoref{fig:random circuit relative distillation time cost increase} show the relative run-time delay and relative distillation time cost increase for each $D$ in $\{1, 2, 4, 8, 16, 32, 64\}$ and $p_{\mathrm{fail}}$ in $\{0.001, 0.002, 0.005, 0.01, 0.02, 0.05\}$, respectively.
The range of $p_{\mathrm{fail}}$ is aligned with the failure probabilities of single-level 15-to-1 distillation protocols in Ref.\cite{Litinski2019magicstate}.
For example, the failure probabilities of the protocols with the physical error probability $10^{-4}$ and $10^{-3}$ are close to $0.001$ and $0.05$, respectively.
As described above, the coupling distance models the locality of couplings: the smaller the coupling distance, the more local the couplings.
The figures show that even a moderate coupling distance and a failure probability incurred a significant additional time cost.

\subsection {Dist-SELECT simulation: scheduling}
\label{subsec:performance-evaluation-scheduling}
We simulated $M$-multiplexed SELECT circuits sharing the Pauli target qubits, which is the core part of the Dist-SELECT circuit.
In the simulation, we assumed we could run at most $P$ controlled Pauli operators in parallel on the Pauli target qubits.
$P$ reflects the routing efficiency that depends on the quantum computer architecture and compiler, but this is the only part of our simulation that takes the routing efficiency into account explicitly.
\autoref{tab:simulation parameters} shows the parameters.

\begin{table}[h!]
\centering
\caption{Dist-SELECT simulation parameters}
\label{tab:simulation parameters}
\begin{tabular}{|l|p{65mm}|}
  \hline
  \textbf{symbol} & \textbf{description}\\
  \hline
  \hline
  $d$ & The code distance of the surface code.\\
  \hline
  $d_{\mathrm{pool}}$ & The code distance for the magic state pool.\\
  \hline
  $\#_{\mathrm{MSD}}$ & The number of magic state factories.\\
  \hline
  $M$ & Dist-SELECT parallelism.\\
  \hline
  $P$ & The number of controlled Pauli operators that can run in parallel on the Pauli target qubits.\\
  \hline
  $N$ & The number of the pauli target qubits.\\
  \hline
  $L$ & The original length of the SELECT operation.\\
  \hline
  $D$ & The distillation cycles used during scheduling.\\
  \hline
  $p_{\mathrm{phys}}$ & The physical error probability. \\
  \hline
\end{tabular}
\end{table}

Scheduling gates in the SELECT circuit is relatively easy because it has an evident critical path and we mostly ignore the topological aspects.
We schedule the operations greedily, with allocating two special resources: magic states and controlled Pauli operators on the Pauli target qubits.
\autoref{alg:schedule} shows the scheduling algorithm.
``Operation" in the algorithm represents a basic operation such as Pauli and measurement, accompanied by dependency information.
Scheduling $M$-multiplexed SELECT circuits with sharing the Pauli target qubits can be implemented by adding small extensions to \autoref{alg:schedule}.

\begin{algorithm}[t!]
\caption{Schedule $operation$}
\label{alg:schedule}
\begin{algorithmic}[1]
  \IF {$operation$ has already been scheduled}
    \STATE \textbf{return}
  \ENDIF
  \STATE $clock \leftarrow 0$
  \FOR {$ancestor$ in the ancestors of $operation$}
    \STATE \textbf{Schedule} $ancestor$.
    \STATE $s \leftarrow ancestor$'s scheduled clock
    \STATE $cycles \leftarrow ancestor$'s time cost
    \STATE $clock \leftarrow {\rm max}(clock, s + cycles)$
  \ENDFOR

  \IF{$operation$ works on the Pauli target qubits}
    \STATE Schedule the operation on the Pauli target qubits, and let $c$ be the scheduled clock.
    \STATE $clock \leftarrow c$
  \ELSIF{$operation$ is a magic state preparation}
    \STATE Request a magic state to factories, and let $c$ be the scheduled clock.
    \STATE $clock \leftarrow c$
  \ENDIF
  \STATE $clock$ is the scheduled clock for $operation$
\end{algorithmic}
\end{algorithm}

Our magic state factory relies heavily on pipelining, and to work efficiently, it needs to store partially completed calculations performed in the previous round of distillation.
On the other hand, in the factory, we use smaller code distances to protect states from errors.
This means states in the factory decay, and we cannot just keep them indefinitely.
Hence, our magic state factories continuously run, and there is no on-demand distillation.
The scheduler expects a magic state factory to generate a magic state every $D$ cycles and the job of the scheduler is to select the
right distillation round for each magic state request.
This is aligned with the execution model discussion in \autoref{subsec:background-parallel-execution}.

\begin{figure}[!t]
  \centering
  \includegraphics[width=8.4cm]{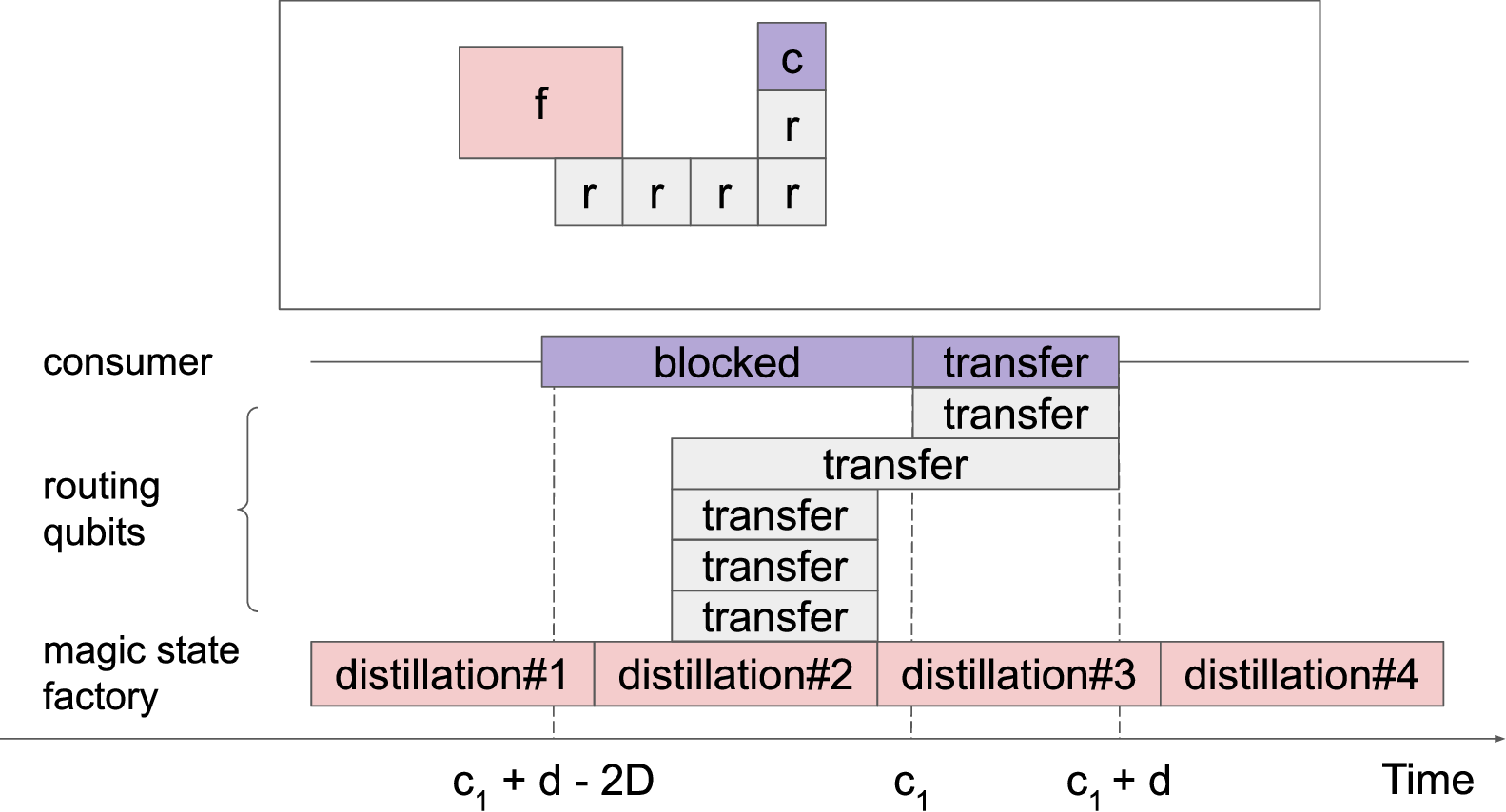}
  \caption{Top figure is a block diagram that depicts a magic state distillation factory (pink), the consumer (purple) and the routing qubits (gray) in the circuit.
           Bottom figure shows the schedule for each component.}
  \label{fig:schedule magic state distillation}
\end{figure}

\autoref{fig:schedule magic state distillation} shows the magic state distillation scheduling process for a scenario with a single factory.
In \autoref{fig:schedule magic state distillation}, the consumer qubit is blocked until $c_1$.
The scheduler looks at the factory's schedule to minimize the duration of the blockage at the consumer qubit of the magic state.
In the example case, distillation \#2 is the best choice, because choosing distillation \#3 would block the consumer.
We call the magic state transfer operation at the consumer (the purple ``transfer'' block in the figure) \textit{magic state preparation}.
In our simulation, routing qubits are treated implicitly and we assume that there are sufficient routing qubits around a distillation factory to avoid congestion.
Hence, despite the fact that the magic state preparation operation and distillation \#3 are performed simultaneously, we assume that there are no conflicts between them and the scheduler can assign distillation \#3 to another magic state request.
We can easily expand this single-factory logic to the multi-factories case, by selecting the best-matching distillation round in all the factories.

If distillation \#2 has already been assigned to another magic state request in \autoref{fig:schedule magic state distillation}, the scheduler needs to look for another one.
Our natural choice would be distillation \#1, but this means that the magic state would be stored in the routing qubits temporally.
Storing states in routing qubits for a long time leads to routing congestion, which our simulation cannot handle, as discussed above.
To avoid this scenario, we set another rule:
The assigned distillation must not precede the magic state preparation operation by more than $2D - d$ cycles.
Hence the scheduler will choose distillation \#3.

\subsection {Dist-SELECT simulation: execution}
\label{subsec:performance-evaluation-execution}

Once the scheduling is complete, we can actually simulate the execution.
The simulation calculates $delay$, delayed cycles against the schedule for each logical qubit (the Pauli target qubits are treated as one qubit).
It processes operations according to the operation dependencies with propagating delay
information.
\autoref{alg:execute} shows the algorithm.

\begin{algorithm}[t!]
\caption{Execute}
\label{alg:execute}
\begin{algorithmic}[1]
  \FOR {$operation$ in all the operations}
    \FOR{i in $0..M$}
      \STATE{$push(queue, (operation, i))$}
    \ENDFOR
  \ENDFOR
  \WHILE{$queue$ is not empty}
    \STATE $(op, index) \leftarrow pop(queue)$
    \IF{$(op, index)$ is complete}
      \STATE \textbf{continue}
    \ELSIF{$(op, index)$ has an incomplete ancestor}
      \STATE \textbf{continue}
    \ENDIF

    \STATE Mark $(op, index)$ as complete.

    \STATE{Let $q$ be the qubit $(op, index)$ is running on.}

    \STATE Propagate delay information based on dependencies.
    \IF {$op$ is a magic state preparation}
      \STATE{Let $d$ be the delay caused by the magic state distillation
             based on the given distribution.}
      \STATE{$delay[q] \leftarrow delay[q] + d$}
    \ENDIF
    \FOR {$(n, i)$ in the descendants of $(op, index)$}
      \STATE{$push(queue, (n, i))$}
    \ENDFOR
  \ENDWHILE
\end{algorithmic}
\end{algorithm}

A delay caused by magic state distillation failures is amplified by the delay propagation process, which is defined by the following criteria.

\begin{itemize}
  \item Propagate a delay between two operations when they have a dependency.
  \item A magic state request needs to wait for all the preceding requests to the associated magic state factory.
\end{itemize}

Particularly, all the controlled Pauli operators on the Pauli target qubits share the Pauli target qubits and have an implicit dependency on each other.
Hence, once a delay is propagated to the Pauli target qubits, the delay affects all the operations running on them thereafter.
This reflects the routing difficulty on the Pauli target qubits.

Let us think about running a magic state preparation operation $o$ at qubit $q$ and how to calculate the delay caused by magic state distillation.
Let $c$ be the scheduled starting timing of $o$, and let $c_{d}$ be the scheduled starting timing of the distillation associated with $o$.
As shown in \autoref{fig:schedule magic state distillation}, $c_{d} \geq c + d - 2D.$
Let $delay$ be the delay on $q$ just before $o$ starts, and $e$ be the first distillation completion that satisfies the following:
\begin{itemize}
  \item Prior magic state requests to the factory have already been completed at the beginning of the distillation.
  \item $e \geq c_{d} + delay + D$.
\end{itemize}

The first condition states that magic state requests are processed in order and
the second condition states that the run-time behavior should not overtake the schedule.
$o$ is completed at ${\rm max}(c + delay + d, e)$ and the new delay on $q$ becomes ${\rm max}(c + delay + d, e) - c - d$.
This behavior can be extended when pooling is enabled, as discussed in \autoref{sec:Mitigations}.

To reduce the simulation time cost, we separate the magic state distillation simulation from the main simulation.
Specifically, we calculate the distribution of the distillation time cost beforehand, and treat each distillation time cost as a random variable with the distribution in the main simulation.

\subsection {Dist-Select simulation: parameter settings}

In this subsection, we describe the parameter settings (see \autoref{tab:simulation parameters} for the list of parameters).
We set some parameters directly from our assumptions, and derived others from them.
We set the problem size $N = 2^{12}$ and $L = 2^{24}$.
$P$ and $M$ control the parallelization level.
We set $P = 32$ and $M = 128$ to simulate highly parallelized circuits.
$p_{\mathrm{phys}}$ is the physical error probability and we chose a realistic value, $10^{-3}$.
We also made the following assumptions:
\begin{itemize}
  \item We want the error probability of the Dist-SELECT circuit to be below 1\%.
  \item We use the \ket{CCZ} distillation protocols in Ref.\cite{Litinski2019magicstate}.
  \item The error probability of a logical qubit per cycle is $p_L(d) \coloneqq 0.1(100p_{\mathrm{phys}})^{\frac{d+1}2}$.
\end{itemize}
To maintain the overall error probability less than 1\%, we wanted the error probabilities of magic states and other components (Clifford operations) to be under 0.5\%.
Because we needed $L$ CCZ states in the circuit, the (15-to-1)$^4_{13,5,5} \times $(8-to-CCZ)$_{23,13,15}$ protocol whose error probability per magic state was $1.8 \times 10^{-10}$ was suitable, resulting in us setting $D = 60$.

To determine the code distance, we needed to roughly estimate the spatial and time cost.
We allocated four logical qubits for each pauli target qubit (similarly to the layout in \autoref{fig:placement-and-routing}) to support the parallel execution of gates, and two logical qubits for each index and ancilla qubit.
Assuming the spatial cost for temporary ancilla qubits was negligible, the total number of logical qubits, except for magic state distillation factories, was $4N + 2\times 2 M\log(\frac{L}{M}) \approx 2.5 \times 10^4$.
The product of $p_L(d)$, spatial cost, and time cost needed to be less than $0.005$; hence, we chose $d = 27$.
We ran the simulation with various $\#_{\mathrm{MSD}}$ values in the range of $[60, 160]$.
60 factories used approximately 5\% and 160 factories used approximately 13\% of the total qubits, both of which were reasonable.
Our experiment showed that more factories would not result in a substantial time cost reduction.

As discussed in \autoref{sec:Mitigations}, $d_{\mathrm{pool}}$, which is the code distance used in the pool, can be smaller than $d$.
Let $E$ be the expected value of the distillation cycles.
Then $d_{\mathrm{pool}}$ must satisfy
\[
    L(3(E+d)p_L(d_{\mathrm{pool}}) + 1.8 \times 10^{-10}) < 0.005.
\]
$d_{\mathrm{pool}} = 23$ is the smallest odd integer satisfying this inequality.
As one factory used 34052 physical qubits, the spatial overhead of the pool was $\frac{3 \times 2d_{\mathrm{pool}}^2}{34052} \approx 9\%$ of the factory size.

\subsection {Dist-SELECT simulation: results}

\begin{figure}[t!]
  \centering
  \includegraphics[width=8.0cm]{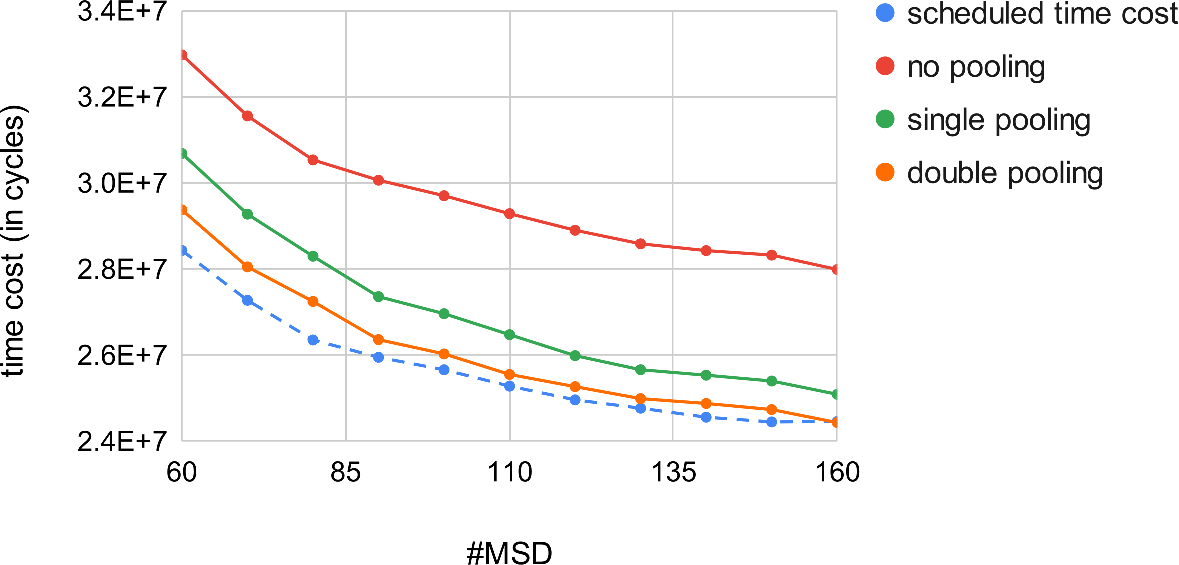}
  \caption{Pooling reduces the run-time delay.
           The blue dotted line shows the scheduled time cost and the other lines represent the total time cost including the run-time delay.}
  \label{fig:pooling and time cost reduction}
\end{figure}

\begin{figure}[!t]
  \centering
  \includegraphics[width=7.0cm]{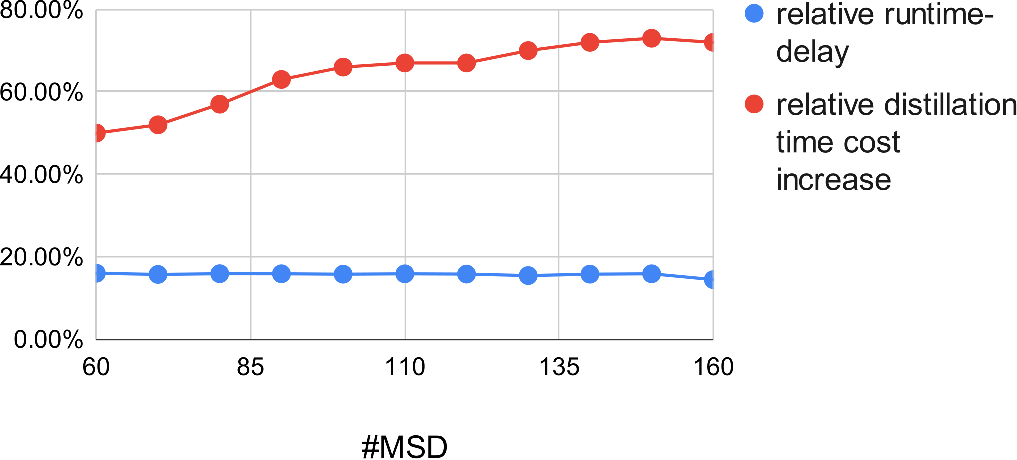}
  \caption{Impact of distillation failures.}
  \label{fig:magic state factories and relative time cost}
\end{figure}

\autoref{fig:pooling and time cost reduction} shows the scheduled time cost (blue dotted line) and total time cost (red solid line), and \autoref{fig:magic state factories and relative time cost} shows the relative run-time delay and relative distillation time cost increase for each $\#_{\mathrm{MSD}}$ value.
As shown in \autoref{fig:magic state factories and relative time cost}, the impact of magic state distillation failures was substantial:
For most $\#_{\mathrm{MSD}}$ values, the relative distillation time cost increase was over 50\%.
This also shows that adding factories reduced the scheduled time cost and hence the total time cost, whereas the relative run-time delay was almost constant.

\begin{figure}[t!]
  \centering
  \includegraphics[width=8.0cm]{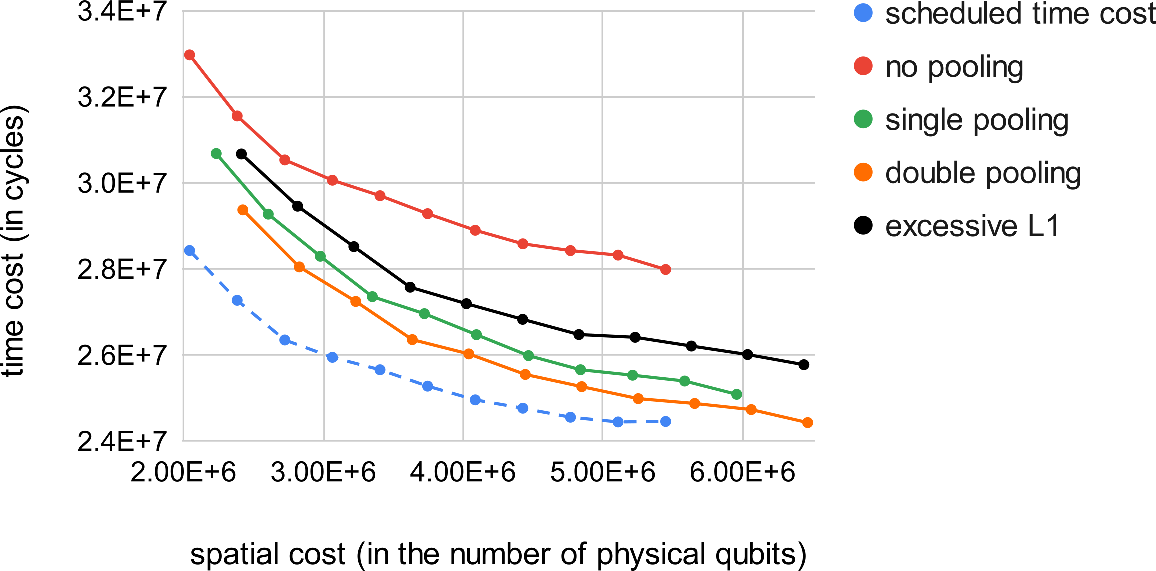}
  \caption{Spatial and time cost trade-off with pooling.}
  \label{fig:pooling result with spatial overhead}
\end{figure}

The impact of pooling on run-time delay is demonstrated in \autoref{fig:pooling and time cost reduction}.
Pooling with single entry (``single pooling") effectively reduced the run-time delay, with reductions ranging from 50\% to 82\%.
Pooling with two entries (``double pooling") further reduced the run-time delay.
To find the best balance between the magic state distillation and consumption speeds for pooling, we ran the simulation with various magic state consumption speeds and selected the value that had the lowest time cost for each $\#_{\mathrm{MSD}}$ setting.
Note that, in the figure, we ignore the spatial overhead of pooling.
We consider it in \autoref{fig:pooling result with spatial overhead}, where the horizontal axis represents the number of physical qubits required for magic state distillation and pooling, instead of $\#_{\mathrm{MSD}}$.
Evidently, the configurations with pooling presented better trade-offs than the configuration without pooling.
Ref.\cite{Beverland2022Assessing} proposed having excessive L1 distillation factories in each L2 distillation factory to mitigate L1 distillation failures, and the black line in \autoref{fig:pooling result with spatial overhead} (``excessive L1") corresponds to the solution.
The figure shows that pooling is better than having excessive L1 factories.

We can also use pooling to reduce the spatial cost.
In \autoref{fig:pooling result with spatial overhead}, $\#_{\mathrm{MSD}} = 150$ without pooling used 5,107,800 qubits and the total time cost was 28,313,114 cycles.
$\#_{\mathrm{MSD}} = 70$ with pooling with two entries used 2,828,000 qubits and the total time cost was 28,038,996 cycles.
Hence, pooling offered a spatial cost reduction of 45\% with a smaller time cost.

The spatial cost of additional magic state factories might be underestimated because when we add a magic state factory, we usually need to add some routing qubits, which is not counted in our assessment.
By contrast, no additional routing qubits are required for pooling.
Hence pooling may be more competitive than \autoref{fig:pooling result with spatial overhead} shows.

\section {Conclusion}
\label{sec:Conclusion}
We indicated that the negative performance impact caused by magic state distillation failures can increase on a parallel-processing quantum computer and we proposed pooling as mitigation.
To see the magnitude of the impact and the usefulness of the mitigation approach, we ran simulations of the random quantum circuit and the Dist-SELECT circuit.

With the random quantum circuit, we observed a relative distillation time cost increase of over 100\% with a modest failure probability and locality.
With the Dist-SELECT circuit, we observed a relative distillation time cost increase of over 50\%.
Hence, we conclude that magic state distillation failures could add a large run-time delay, particularly when the circuit is highly parallelized, and system designers need to pay attention to such failures.
Although the actual numbers depend on the computer architecture and the problem, we believe that the overall conclusion is robust.

With the Dist-SELECT circuit, we observed that pooling reduced much of the run-time delay, with a small additional distillation spatial cost.
Pooling offered better space-time trade-offs compared with configurations without it.
We also observed a 45\% distillation spatial cost reduction without increasing the time cost.
Hence, we conclude that pooling is useful to mitigate the negative performance impact of magic state distillation failures.
Moreover, we observed that pooling is a better approach compared to having excessive L1 distillation factories in each L2 distillation factory~\cite{Beverland2022Assessing}.

The scheduler described in \autoref{sec:Performance evaluation} is simple.
More sophisticated schedulers should be able to exploit parallelism more to shorten the program execution time.
For example, commuting operations on the same qubits can run in parallel with extra space overhead\cite{Litinski2019GameOfSurfaceCodes}.
With such sophisticated schedulers, the effect of magic state distillation failures can increase more; hence, the importance of mitigation to the effect of the failures.

We have focused on magic state distillation simply because it is a frequently used RUS operation.
We believe that any frequently used RUS operations could lead to similar results, whereas mitigation may be different from those proposed in this paper.
For example, \cite{Akahoshi2023Partially} used RUS operations massively.
State injection used by magic state distillation is another example of probabilistic procedure \cite{Fowler2019Low,Gidney2023Cleaner,Shraddha2022}.
Although they are not distillation protocols, their failures may have similar effects on the computation time.


\newpage


\bibliographystyle{plain}
\bibliography{refs}

\begin{thebibliography}{10}

\bibitem{Abrams1999Quantum}
Daniel~S. Abrams and Seth Lloyd.
\newblock Quantum algorithm providing exponential speed increase for finding
  eigenvalues and eigenvectors.
\newblock {\em Phys. Rev. Lett.}, 83:5162--5165, Dec 1999.

\bibitem{Akahoshi2023Partially}
Yutaro Akahoshi, Kazunori Maruyama, Hirotaka Oshima, Shintaro Sato, and Keisuke
  Fujii.
\newblock Partially fault-tolerant quantum computing architecture with
  error-corrected clifford gates and space-time efficient analog rotations,
  2023.

\bibitem{Amdahl1967}
Gene~M. Amdahl.
\newblock Validity of the single processor approach to achieving large scale
  computing capabilities.
\newblock In {\em Proceedings of the April 18-20, 1967, Spring Joint Computer
  Conference}, AFIPS '67 (Spring), page 483–485, New York, NY, USA, 1967.
  Association for Computing Machinery.

\bibitem{Amy2014Polynomial}
Matthew Amy, Dmitri Maslov, and Michele Mosca.
\newblock Polynomial-time t-depth optimization of clifford+t circuits via
  matroid partitioning.
\newblock {\em IEEE Transactions on Computer-Aided Design of Integrated
  Circuits and Systems}, 33(10):1476--1489, 2014.

\bibitem{Amy2013MeetInTheMiddle}
Matthew Amy, Dmitri Maslov, Michele Mosca, and Martin Roetteler.
\newblock A meet-in-the-middle algorithm for fast synthesis of depth-optimal
  quantum circuits.
\newblock {\em IEEE Transactions on Computer-Aided Design of Integrated
  Circuits and Systems}, 32(6):818--830, 2013.

\bibitem{Anderson2014FaultTolerantConversion}
Jonas~T. Anderson, Guillaume Duclos-Cianci, and David Poulin.
\newblock Fault-tolerant conversion between the steane and reed-muller quantum
  codes.
\newblock {\em Phys. Rev. Lett.}, 113:080501, Aug 2014.

\bibitem{Arute2019QuantumSupremacy}
Frank Arute, Kunal Arya, Ryan Babbush, Dave Bacon, Joseph Bardin, Rami Barends,
  Rupak Biswas, Sergio Boixo, Fernando Brandao, David Buell, Brian Burkett,
  Yu~Chen, Jimmy Chen, Ben Chiaro, Roberto Collins, William Courtney, Andrew
  Dunsworth, Edward Farhi, Brooks Foxen, Austin Fowler, Craig~Michael Gidney,
  Marissa Giustina, Rob Graff, Keith Guerin, Steve Habegger, Matthew Harrigan,
  Michael Hartmann, Alan Ho, Markus~Rudolf Hoffmann, Trent Huang, Travis
  Humble, Sergei Isakov, Evan Jeffrey, Zhang Jiang, Dvir Kafri, Kostyantyn
  Kechedzhi, Julian Kelly, Paul Klimov, Sergey Knysh, Alexander Korotkov, Fedor
  Kostritsa, Dave Landhuis, Mike Lindmark, Erik Lucero, Dmitry Lyakh, Salvatore
  Mandrà, Jarrod~Ryan McClean, Matthew McEwen, Anthony Megrant, Xiao Mi,
  Kristel Michielsen, Masoud Mohseni, Josh Mutus, Ofer Naaman, Matthew Neeley,
  Charles Neill, Murphy~Yuezhen Niu, Eric Ostby, Andre Petukhov, John Platt,
  Chris Quintana, Eleanor~G. Rieffel, Pedram Roushan, Nicholas Rubin, Daniel
  Sank, Kevin~J. Satzinger, Vadim Smelyanskiy, Kevin~Jeffery Sung, Matt
  Trevithick, Amit Vainsencher, Benjamin Villalonga, Ted White, Z.~Jamie Yao,
  Ping Yeh, Adam Zalcman, Hartmut Neven, and John Martinis.
\newblock Quantum supremacy using a programmable superconducting processor.
\newblock {\em Nature}, 574:505–510, 2019.

\bibitem{Aspuru-Guzik2005SimulatedQuantumComputation}
Alán Aspuru-Guzik, Anthony~D. Dutoi, Peter~J. Love, and Martin Head-Gordon.
\newblock Simulated quantum computation of molecular energies.
\newblock {\em Science}, 309(5741):1704--1707, 2005.

\bibitem{Babbush2018}
Ryan Babbush, Craig Gidney, Dominic~W. Berry, Nathan Wiebe, Jarrod McClean,
  Alexandru Paler, Austin Fowler, and Hartmut Neven.
\newblock Encoding electronic spectra in quantum circuits with linear t
  complexity.
\newblock {\em Phys. Rev. X}, 8:041015, Oct 2018.

\bibitem{Beverland2022EdgeDisjointPath}
Michael Beverland, Vadym Kliuchnikov, and Eddie Schoute.
\newblock Surface code compilation via edge-disjoint paths.
\newblock {\em PRX Quantum}, 3:020342, May 2022.

\bibitem{Beverland2021CostOfUniversality}
Michael~E. Beverland, Aleksander Kubica, and Krysta~M. Svore.
\newblock Cost of universality: A comparative study of the overhead of state
  distillation and code switching with color codes.
\newblock {\em PRX Quantum}, 2:020341, Jun 2021.

\bibitem{Beverland2022Assessing}
Michael~E. Beverland, Prakash Murali, Matthias Troyer, Krysta~M. Svore, Torsten
  Hoefler, Vadym Kliuchnikov, Guang~Hao Low, Mathias Soeken, Aarthi Sundaram,
  and Alexander Vaschillo.
\newblock Assessing requirements to scale to practical quantum advantage, 2022.

\bibitem{Bombín2024FaultTolerant}
H\'ector Bomb\'{\i}n, Mihir Pant, Sam Roberts, and Karthik~I. Seetharam.
\newblock Fault-tolerant postselection for low-overhead magic state
  preparation.
\newblock {\em PRX Quantum}, 5:010302, Jan 2024.

\bibitem{Bombín2015GaugeColorCodes}
Héctor Bombín.
\newblock Gauge color codes: optimal transversal gates and gauge fixing in
  topological stabilizer codes.
\newblock {\em New Journal of Physics}, 17(8):083002, aug 2015.

\bibitem{Bombín2016DimentionalJump}
Héctor Bombín.
\newblock Dimensional jump in quantum error correction.
\newblock {\em New Journal of Physics}, 18(4):043038, apr 2016.

\bibitem{Bouland2018OnTheComplexity}
Adam Bouland, Bill Fefferman, Chinmay Nirkhe, and Umesh~V. Vazirani.
\newblock On the complexity and verification of quantum random circuit
  sampling.
\newblock {\em Nature Physics}, 15:159--163, 2018.

\bibitem{Bravyi1998}
S.~B. Bravyi and A.~Yu. Kitaev.
\newblock Quantum codes on a lattice with boundary, 1998.

\bibitem{Bravyi2005}
Sergey Bravyi and Alexei Kitaev.
\newblock Universal quantum computation with ideal clifford gates and noisy
  ancillas.
\newblock {\em Phys. Rev. A}, 71:022316, Feb 2005.

\bibitem{Campbell2017Unified}
Earl~T. Campbell and Mark Howard.
\newblock Unified framework for magic state distillation and multiqubit gate
  synthesis with reduced resource cost.
\newblock {\em Phys. Rev. A}, 95:022316, Feb 2017.

\bibitem{Eastin2009RestrictionsOnTransversal}
Bryan Eastin and Emanuel Knill.
\newblock Restrictions on transversal encoded quantum gate sets.
\newblock {\em Phys. Rev. Lett.}, 102:110502, Mar 2009.

\bibitem{Fowler2019Low}
Austin~G. Fowler and Craig Gidney.
\newblock Low overhead quantum computation using lattice surgery, 2019.

\bibitem{Gidney2023Cleaner}
Craig Gidney.
\newblock Cleaner magic states with hook injection, 2023.

\bibitem{Gidney2021howtofactorbit}
Craig Gidney and Martin Eker{\aa{}}.
\newblock How to factor 2048 bit {RSA} integers in 8 hours using 20 million
  noisy qubits.
\newblock {\em {Quantum}}, 5:433, April 2021.

\bibitem{Gidney2019EfficientMagicState}
Craig Gidney and Austin~G. Fowler.
\newblock Efficient magic state factories with a catalyzed {$|CCZ\rangle$} to
  {$2|T\rangle$} transformation.
\newblock {\em {Quantum}}, 3:135, April 2019.

\bibitem{Herr2017LatticeSurgery}
Daniel Herr, Franco Nori, and Simon~J Devitt.
\newblock Lattice surgery translation for quantum computation.
\newblock {\em New Journal of Physics}, 19(1):013034, jan 2017.

\bibitem{Horsman2012}
Dominic Horsman, Austin~G Fowler, Simon Devitt, and Rodney~Van Meter.
\newblock Surface code quantum computing by lattice surgery.
\newblock {\em New Journal of Physics}, 14(12):123011, dec 2012.

\bibitem{Kitaev1995Quantum}
A.~Yu. Kitaev.
\newblock Quantum measurements and the abelian stabilizer problem, 1995.

\bibitem{Kitaev2003}
A.Yu. Kitaev.
\newblock Fault-tolerant quantum computation by anyons.
\newblock {\em Annals of Physics}, 303(1):2--30, 2003.

\bibitem{Litinski2019GameOfSurfaceCodes}
Daniel Litinski.
\newblock A {G}ame of {S}urface {C}odes: {L}arge-{S}cale {Q}uantum {C}omputing
  with {L}attice {S}urgery.
\newblock {\em {Quantum}}, 3:128, March 2019.

\bibitem{Litinski2019magicstate}
Daniel Litinski.
\newblock Magic {S}tate {D}istillation: {N}ot as {C}ostly as {Y}ou {T}hink.
\newblock {\em {Quantum}}, 3:205, December 2019.

\bibitem{Martyn2021GrandUnification}
John~M. Martyn, Zane~M. Rossi, Andrew~K. Tan, and Isaac~L. Chuang.
\newblock Grand unification of quantum algorithms.
\newblock {\em PRX Quantum}, 2:040203, Dec 2021.

\bibitem{McEwen2023RelaxingHardware}
Matt McEwen, Dave Bacon, and Craig Gidney.
\newblock Relaxing {H}ardware {R}equirements for {S}urface {C}ode {C}ircuits
  using {T}ime-dynamics.
\newblock {\em {Quantum}}, 7:1172, November 2023.

\bibitem{Paetznick2013FaultTolerant}
Adam Paetznick and Ben~W. Reichardt.
\newblock Universal fault-tolerant quantum computation with only transversal
  gates and error correction.
\newblock {\em Phys. Rev. Lett.}, 111:090505, Aug 2013.

\bibitem{Paetznick2014Repeat}
Adam Paetznick and Krysta~M. Svore.
\newblock Repeat-until-success: Non-deterministic decomposition of single-qubit
  unitaries.
\newblock {\em Quantum Info. Comput.}, 14(15–16):1277–1301, nov 2014.

\bibitem{Riesebos2017}
L.~Riesebos, X.~Fu, S.~Varsamopoulos, C.~G. Almudever, and K.~Bertels.
\newblock Pauli frames for quantum computer architectures.
\newblock In {\em Proceedings of the 54th Annual Design Automation Conference
  2017}, DAC '17, New York, NY, USA, 2017. Association for Computing Machinery.

\bibitem{Selinger2013Quantum}
Peter Selinger.
\newblock Quantum circuits of $t$-depth one.
\newblock {\em Phys. Rev. A}, 87:042302, Apr 2013.

\bibitem{Shor1994Factoring}
Peter~W. Shor.
\newblock Algorithms for quantum computation: discrete logarithms and
  factoring.
\newblock {\em Proceedings 35th Annual Symposium on Foundations of Computer
  Science}, pages 124--134, 1994.

\bibitem{Shraddha2022}
Shraddha Singh, Andrew~S. Darmawan, Benjamin~J. Brown, and Shruti Puri.
\newblock High-fidelity magic-state preparation with a biased-noise
  architecture.
\newblock {\em Phys. Rev. A}, 105:052410, May 2022.

\bibitem{Yoshioka2022Hunting}
Nobuyuki Yoshioka, Tsuyoshi Okubo, Yasunari Suzuki, Yuki Koizumi, and Wataru
  Mizukami.
\newblock Hunting for quantum-classical crossover in condensed matter problems,
  2022.

\bibitem{Zhang2021}
Mengyu Zhang, Lei Xie, Zhenxing Zhang, Qiaonian Yu, Guanglei Xi, Hualiang
  Zhang, Fuming Liu, Yarui Zheng, Yicong Zheng, and Shengyu Zhang.
\newblock Exploiting different levels of parallelism in the quantum control
  microarchitecture for superconducting qubits.
\newblock In {\em MICRO-54: 54th Annual IEEE/ACM International Symposium on
  Microarchitecture}, MICRO '21, page 898–911, New York, NY, USA, 2021.
  Association for Computing Machinery.

\end{thebibliography}

\end{document}